%% file: elsarticle-template.tex
\documentclass[review]{elsarticle}

\usepackage{hyperref}

\journal{Journal of \LaTeX\ Templates}

\usepackage[english]{babel}
\usepackage[utf8x]{inputenc}
\usepackage[T1]{fontenc}
\usepackage{amsmath}
\usepackage{mathtools}
\usepackage{tkz-graph}
\usepackage{multicol}
\usepackage{fullpage}

\usepackage{amsmath}
\usepackage{multicol}
\usepackage{amsfonts}
\usepackage{graphicx}
\usepackage{paralist}
\usepackage[colorinlistoftodos]{todonotes}
\usepackage{rotating}
\usepackage{multirow}
\usepackage{algorithm}
\usepackage{enumitem}
\usepackage{algorithmic}
\setlist[itemize]{noitemsep, topsep=0pt}
\setlist[enumerate]{noitemsep, topsep=0pt}

\newdefinition{rmk}{Remark} 
\newproof{pf}{Proof}
\newproof{pot}{Proof of Theorem \ref{thm2}}
\begin{document}

\begin{frontmatter}

\title{Scalable Optimal Deployment in the Cloud of \\Component-based Applications using Optimization Modulo Theory, Mathematical Programming and Symmetry Breaking\tnoteref{mytitlenote}}
\tnotetext[mytitlenote]{This work was partially supported by a grant of the Romanian National Authority for Scientific Research and Innovation, CNCS/CCCDI - UEFISCDI, project number PN-III-P2-2.1-PED-2016-0550, within PNCDI III.}
\author[uvt]{M\u{a}d\u{a}lina Era\c{s}cu\corref{cor1}}
\ead{madalina.erascu@e-uvt.ro}
\author[uvt]{Flavia Micota}
\ead{flavia.micota@e-uvt.ro}
\author[uvt]{Daniela Zaharie}
\ead{daniela.zaharie@e-uvt.ro}

\cortext[cor1]{Corresponding author}

\address[uvt]{Faculty of Mathematics and Informatics, West University of Timi\c{s}oara, Romania}

\begin{abstract}
The problem of \emph{Cloud resource provisioning for component-based applications} consists in the allocation of virtual machines (VMs) offers from various Cloud Providers to a set of applications such that the constraints induced by the interactions between components and by the components hardware/software requirements are satisfied and the performance objectives are optimized (e.g. costs are minimized). It can be formulated as a constraint optimization problem, hence, in principle the optimization can be carried out automatically. In the case the set of VM offers is large (several hundreds), the computational requirement is huge, making the automatic optimization practically impossible with the current general optimization modulo theory (OMT) and mathematical programming (MP) tools.
We overcame the difficulty by methodologically analyzing the particularities of the problem with the aim of identifying search space reduction methods. These are methods exploiting:
\begin{inparaenum}[\itshape (i)\upshape]
\item the \emph{symmetries} of the general Cloud deployment problem, 
\item the \emph{graph representation} associated to the structural constraints specific to each particular application, and
\item their combination.
\end{inparaenum}
An extensive experimental analysis has been conducted on four classes of real-world problems, using six symmetry breaking strategies and two types of optimization solvers.

As a result, the combination of a variable reduction strategy with a column-wise symmetry breaker leads to a scalable deployment solution, when OMT is used to solve the resulting optimization problem.

\end{abstract}


\end{frontmatter}

\input{introduction}
\input{casestudies}
\input{problem}
\input{solution}
\input{experimentalanalysis}

\input{relatedwork}
\input{conclusions}
\bibliography{mybibfile}
\end{document}

%% file: introduction.tex
\section{Introduction}\label{sec:Introduction}
Efficient resource management in the context of deploying component-based software applications in the Cloud means deciding which virtual machines (VMs) to acquire from the Cloud Providers (CP) and how to place the software components on them in such a way that the functional architecture is preserved and the deployment cost is minimized. Automated Cloud resource provisioning requires solving a selection and an assignment problem, i.e. which VMs should be leased from CPs and how should the components be assigned to them in such a way that the cost is minimized. This is related to the \emph{bin-packing problem}, a fundamental problem in \emph{combinatorial optimization}, which arises in many challenging problems from diverse application areas. Due to the importance of the bin-packing problem, there has been extensive research on developing mathematical formalisms, efficient algorithms, software systems, and applications (just to name a few:~ \cite{DBLP:journals/toms/MartelloPVBK07,doi:10.1137/0207001,FernandezdelaVega1981,Beloglazov:2010:EER:1844765.1845139,DBLP:journals/eor/Carvalho02}). It can be formulated as follows~ \cite{Korte:2012:COT:2190621}:   given a set of bins $V_1, V_2, ..., V_M$ with the same size $V$ and a list of $N$ items with sizes $\emph{R}_1, ..., \emph{R}_N$ find:
\begin{inparaenum}[\itshape (i)\upshape] 
	\item the minimum number $m$ of bins, and
	\item a $m$-partition $V_1 \cup ... \cup V_m$ of the set $\{1,...,N\}$,
\end{inparaenum} such that the objects assigned to the bins do not exceed their capacity ($\sum\limits_{i \in V_k} R_i \le V, \ \forall k = \overline{1,m}$). The problem can be formulated as a constrained optimization problem (COP) as follows: 
\hrule
\[
\begin{tabular}{rl}

\emph{Minimize} $m=\sum\limits_{k=1}^M v_k$ \emph{subject to} & $m \ge 1, \ a_{ik} \in \{0, 1\}, \ v_{k} \in \{0, 1\}$\\
&$\sum\limits_{i=1}^N \emph{R}_i a_{ik} \le V v_k$, \\
&$\sum\limits_{k=1}^M a_{ik} =1, \quad  i=\overline{1,N}$\\
\emph{where} &$v_k = 1$ if bin $k$ is used and $a_{ik}=1$ if item $i$ is placed in bin $k$.
\end{tabular}
\]
\hrule

\smallskip

In a recent project\footnote{\url{https://merascu.github.io/links/MANeUveR.html}}, we studied the problem of \emph{Cloud resource provisioning for component-based applications}. It consists in the allocation of virtual machines (VMs) offers from various Cloud Providers (CPs), to a set of applications such that the constraints induced by the interactions between components and by the components hardware/software requirements are satisfied and the performance objectives are optimized (e.g. costs are minimized). 

The problem is \emph{similar to the bin-packing problem}, however:
\begin{enumerate}
	\item bins (VMs) can have different capacity, which depends on the VMs offers;
	\item the placement of items (components) in bins is limited not only by the capacity constraints, but also by the constraints induced by the components interactions;
	\item the number of items is not known a priori (for component-based applications, several instances of a component can be deployed, depending on specific constraints on the number of instances);
	\item the smallest cost (optimality criteria) is not necessarily obtained by minimizing the number of bins.
\end{enumerate} 

It can be formulated as a constraint optimization problem (COP) and solved, in principle, by state-of-the-art \emph{mathematical programming (MP)} and \emph{optimization modulo theories (OMT) tools}. While the application of MP techniques for solving COP has a long tradition, the usage of OMT is recent. Our motivation for using the OMT approach lies in the tremendous advances of methods and tools in this domain in the last decade. Applications in artificial intelligence and formal methods for hardware and software development have greatly benefited from these \cite{8602994,10.1145/2597809.2597817,DBLP:conf/tacas/NadelR16}. The performance of these tools is highly dependent on the way the problem is formalized as this determines the size of the search space. In the case when the number of VMs offers is large, a naive encoding which does not exploit the symmetries of the underlying problem leads to a huge search space making the optimization problem intractable. We overcame this issue by reducing the search space by:
\begin{enumerate}
	\item systematically analyzing the symmetries which appear in the context of Cloud deployment applications;
	\item design and integrate with state-of-the-art MP (CPLEX \cite{2016_CPLEX_usermanual}) and OMT (Z3 \cite{DBLP:conf/tacas/BjornerPF15}) tools static symmetry breakers for speeding-up the solution process.
\end{enumerate}

As a result the scalability of the used optimizers increased, most notable, by at least $2$ orders of magnitude in the case of the OMT solver.

This paper extends our previous work \cite{DBLP:conf/iccp2/MicotaEZ18,LPAR-IWIL2018:Influence_of_Variables_Encoding,Erascu_SMT_2019} in the following aspects:
\begin{enumerate}
	\item we formalize the Cloud deployment problem (Section~\ref{sec:problem}) by abstracting away the particularities of several realistic case studies (Section~\ref{sec:casestudies});
	\item we propose a methodology analyzing the particularities of the problem with the aim of identifying search space reduction methods; these are methods exploiting the \emph{symmetries} of the general Cloud deployment problem, respectively methods utilizing the \emph{graph representation (cliques)} of each application (Section~\ref{sec:solution});
	\item we assess and compare the performance of two tools based on different theoretical background, namely mathematical programming (CPLEX~\cite{2016_CPLEX_usermanual}) and computational logic (Z3~\cite{10.1007/978-3-540-78800-3_24}); we identified limits in their scalability and applied search space reduction methods aiming to improve their performance (Section~\ref{sec:experimentalanalysis}).
\end{enumerate}



%% file: casestudies.tex
\section{Case Studies}\label{sec:casestudies}
The case studies introduced in this section exemplify the following aspects:
\begin{inparaenum}[\itshape (i)\upshape]
	\item different component characteristics and the rich interactions type in between;
	\item the kind of linear constraints used to express these interactions (see exemplifications in Section~\ref{sec:problem});
	\item the kind of solution we are searching for  (see Section~\ref{sec:solution}).
\end{inparaenum}

\subsection{Secure Web Container}\label{sec:SecWebContainer} 
\begin{figure}[h!]
	\centering
	\includegraphics[width=0.7\textwidth]{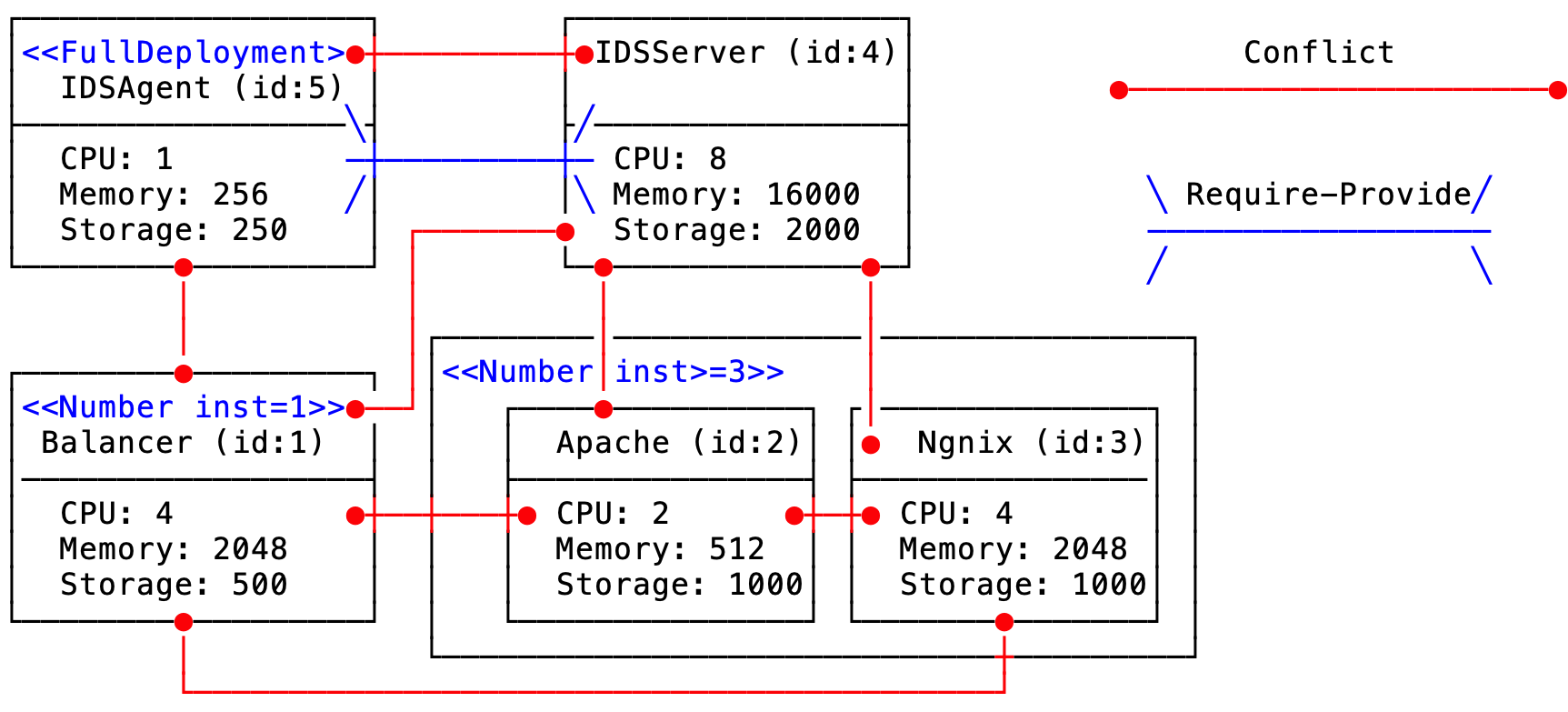}
	\caption{Secure Web Container Application}
	\label{fig:SecureWebContainer}
\end{figure}

The \emph{Secure Web Container} \cite{DBLP:journals/tsc/CasolaBEMR17} (Figure~\ref{fig:SecureWebContainer}) is a service which provides:
\begin{inparaenum}[\itshape (i)\upshape]
	\item \emph{resilience} to attacks and failures, by introducing redundancy and diversity techniques, and
	\item protection from unauthorized and potentially dangerous accesses, by integrating proper \emph{intrusion detection} tools.
\end{inparaenum}
Resilience can be implemented by a set of different Web Container components and a Balancer component, which is responsible for dispatching web requests to the active web containers to ensure load balancing. In the simplest scenario, there are two Web Containers (e.g. Apache Tomcat\footnote{\url{http://tomcat.apache.org}} and Nginx). Intrusion detection is ensured by the generation of intrusion detection reports with a certain frequency. It was implemented by deploying an IDSAgent, to be installed on the resources to be protected, and an IDSServer, which collects data gathered by the IDSAgents and performs the detection activities.

\noindent The constraints between application components are as follows.
\begin{itemize}
	\item For Web resilience:
	\begin{inparaenum}[\itshape (i)\upshape]
		\item Any two of the Balancer, Apache and Nginx components cannot be deployed on the same machine (\emph{\textbf{Conflict}} constraint);
		\item Exactly one Balancer component has to be instantiated (\emph{\textbf{Deployment with bounded number of instances}} constraint, in particular \emph{\textbf{equal bound}}).
		\item The total number of instances for Apache and Nginx components must be at least $3$ (level of redundancy) (\emph{\textbf{Deployment with bounded number of instances}} constraint, in particular \emph{\textbf{lower bound}}).
	\end{inparaenum}
	\item For Web intrusion detection:
	\begin{inparaenum}[\itshape (i)\upshape]
		\item the IDSServer component needs exclusive use of machines (\emph{\textbf{Conflict}} constraint).
		\item There must be an IDSServer component additional instance every 10 IDSAgent component instances (\emph{\textbf{Require-Provide}} constraint).
		\item One instance of IDSAgent must be allocated on every acquired machine except where an IDSServer or a Balancer are deployed (\emph{\textbf{Full Deployment}} constraint).
	\end{inparaenum}
\end{itemize}

We want to deploy this application in the Cloud with the minimal cost. There are multiple Cloud Providers that offer infrastructure services (virtual machines) in multiple heterogeneous configurations, including Amazon\footnote{https://aws.amazon.com}, Google
Cloud\footnote{https://cloud.google.com/compute/}, Microsoft Azure\footnote{https://azure.microsoft.com/en-us/}. In fact, the Crawler Engine we implemented~\cite{DBLP:conf/synasc/ErascuIM18} gathered several hundreds of virtual machines offers having different types, i.e distinct hardware configurations (e.g. number of CPUs, memory, storage) and prices. Therefore the method used to solve the constraint optimization problem should be scalable with respect to the number of VM offers.
\subsection{Secure Billing Email Service}\label{sec:SecureBillingEmailService}
\begin{figure}[h!]
	\centering
	\includegraphics[width=0.55\textwidth]{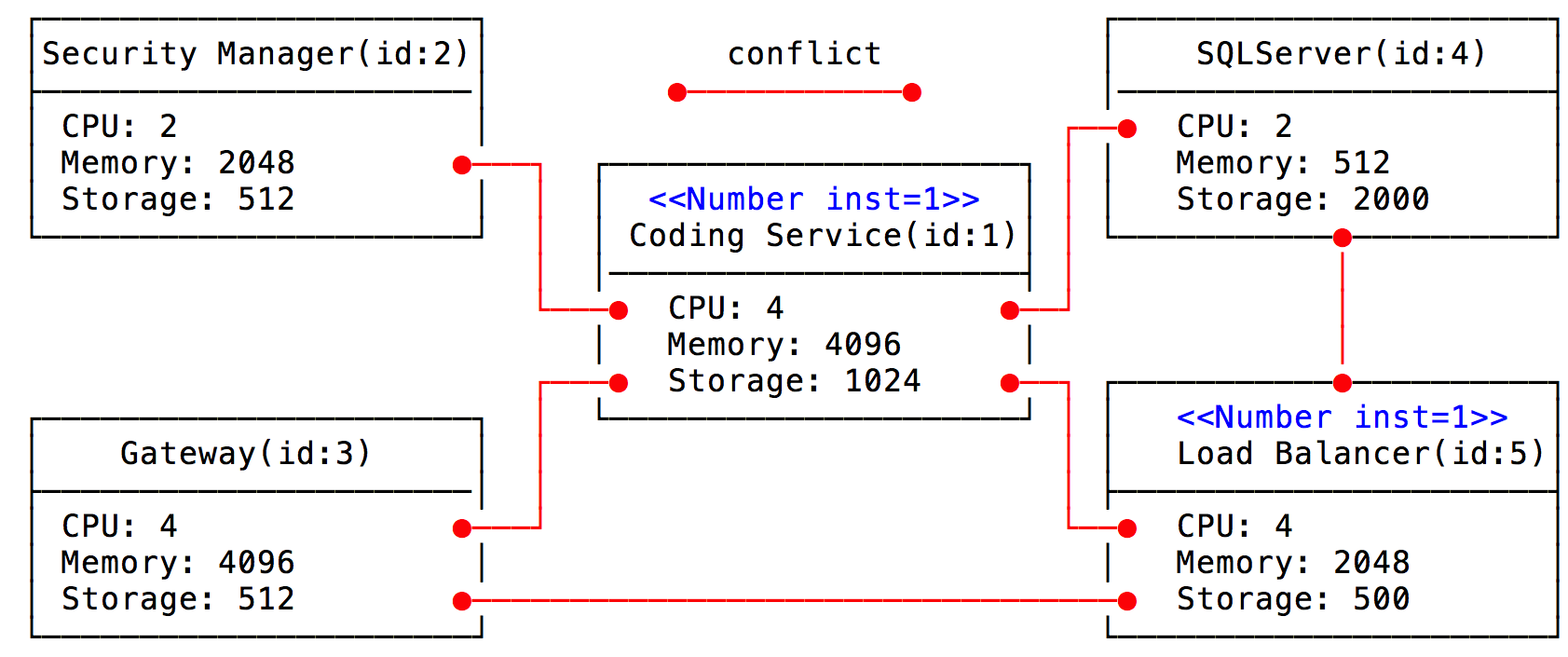}
	\caption{Secure Billing Email Service}
	\label{fig:SecureBillingEmail}
\end{figure}

In the context of a web application ensuring a secure billing email service (Figure~\ref{fig:SecureBillingEmail}) we consider an architecture consisting of $5$ components: 
\begin{inparaenum}[\itshape (i)\upshape]
	\item a coding service ($C_1$), 
	\item a software manager of the user rights and privileges ($C_2$), 
	\item a gateway component ($C_3$), 
	\item an SQL server ($C_4$) and 
	\item a load balancer ($C_5$).
\end{inparaenum} 
Component $C_1$ should use exclusively a virtual machine, thus it can be considered in \emph{\textbf{conflict}} with all the other components. In such a case the original optimization problem can be decomposed in two subproblems, one corresponding to component $C_1$ and the other one corresponding to the other $4$ components. The first problem is trivial: find the VM with the smallest price which satisfies the hardware requirements of component $C_1$. 

The load balancing component should not be placed on the same machine as the gateway component and the SQL server (\emph{\textbf{Conflict}} constraint). On the other hand, only one instance of components $C_1$ and $C_5$ should be deployed while the other three components could have a larger number of instances placed on different virtual machines (\emph{\textbf{Deployment with bounded number of instances}} constraint, in particular \emph{\textbf{equal bound}}).
\subsection{Wordpress}\label{sec:Wordpress}
\begin{figure}[h!]
	\centering
	\includegraphics[width=0.55\textwidth]{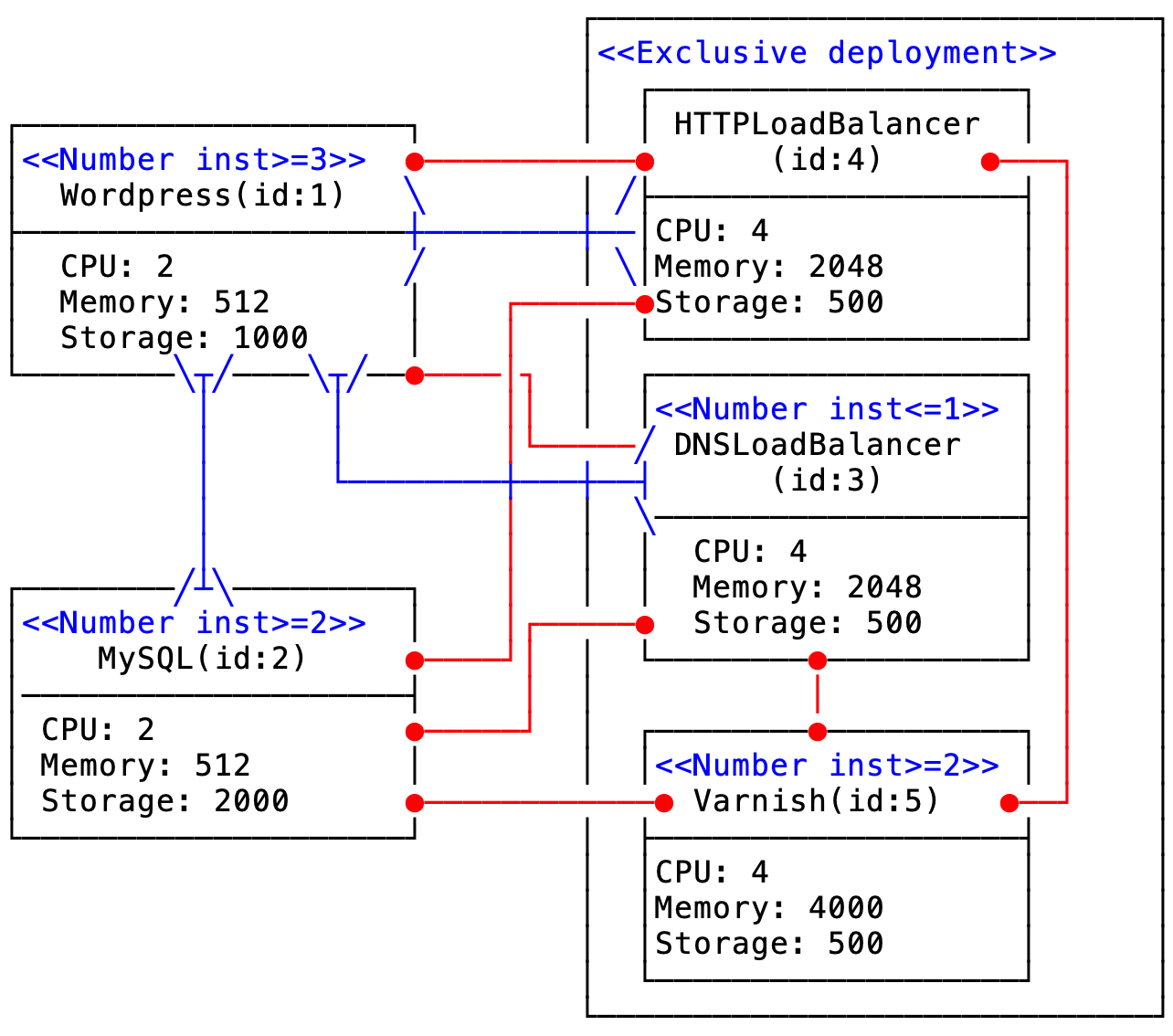}
	\caption{Wordpress Application}
	\label{fig:Wordpress}
\end{figure}

\emph{Wordpress} open-source application is frequently used in creating websites, blogs and applications. We chose it in order to compare our approach to Zephyrus and Zephyrus2 deployment tools~\cite{DBLP:conf/kbse/CosmoLTZZEA14,DBLP:conf/setta/AbrahamCJKM16}. In \cite{DBLP:conf/kbse/CosmoLTZZEA14,DBLP:conf/setta/AbrahamCJKM16}, the authors present a high-load and fault tolerant Wordpress (Figure~\ref{fig:Wordpress}) deployment scenario. The two characteristics are ensured by load balancing. One possibility is to balance load at the DNS level using servers like Bind\footnote{\url{https://www.isc.org/downloads/bind/}}: multiple DNS requests to resolve the website name will result in different IPs from a given pool of machines, on each of which a separate Wordpress instance is running. Alternatively one can use as website entry point an HTTP reverse proxy capable of load balancing (and caching, for added benefit) such as Varnish. In both cases, Wordpress instances will need to be configured to connect to the same MySQL database, to avoid delivering inconsistent results to users. Also, having redundancy and balancing at the front-end level, one usually expects to have them also at the Database Management System (DBMS) level. One way to achieve that is to use a MySQL cluster, and configure the Wordpress instances with multiple entry points to it.
 In the deployment scenario considered by us, the following constraints must be fulfilled:
\begin{inparaenum}[\itshape (i)\upshape]
	\item DNSLoadBalancer requires at least one instance of Wordpress and DNSLoadBalancer can serve at most 7 Wordpress instances (\emph{\textbf{Require-Provide}} constraint).
	\item HTTPLoadBalancer requires at least one Wordpress instance and HTTPLoadBalancer can serve at most 3 Wordpress instances (\emph{\textbf{Require-Provide}} constraint).
	\item Wordpress requires at least three instances of MySQL and MySQL can serve at most 2 Wordpress (\emph{\textbf{\textbf{Require-Provide}}} constraint).
	\item Only one type of Balancer must be deployed; the Balancer components are HTTPLoadBalancer, DNSLoadBalancer and Varnish (\emph{\textbf{Exclusive deployment}} constraint).
	\item Since Varnish exhibits load balancing features, it should not be deployed on the same VM as with another type of Balancer (\emph{\textbf{Conflict}} constraint). Moreover, Varnish and MySQL should not be deployed on the same VM because it is best practice to isolate the DBMS level of an application (\emph{\textbf{Conflict}} constraint).
	\item If HTTPLoadBalancer is deployed, then at least 2 instances of Varnish must be deployed too  (\emph{\textbf{Deployment with bounded number of instances}} constraint, in particular \emph{\textbf{lower bound}}).
	\item At least 2 different entry points to the MySQL cluster (\emph{\textbf{Deployment with bounded number of instance}} constraint, in particular \emph{\textbf{lower bound}}).
	\item No more than 1 DNS server deployed in the administrative domain (\emph{\textbf{Deployment with bounded number of instances}} constraint, in particular \emph{\textbf{upper bound}}).
	\item Balancer components must be placed on a single VM, so they are considered to be in \emph{\textbf{conflict}} with all the other components.
\end{inparaenum}

\subsection{Oryx2}\label{sec:oryx2}
\begin{figure}[h!]
	\centering
	\includegraphics[width=0.65\textwidth]{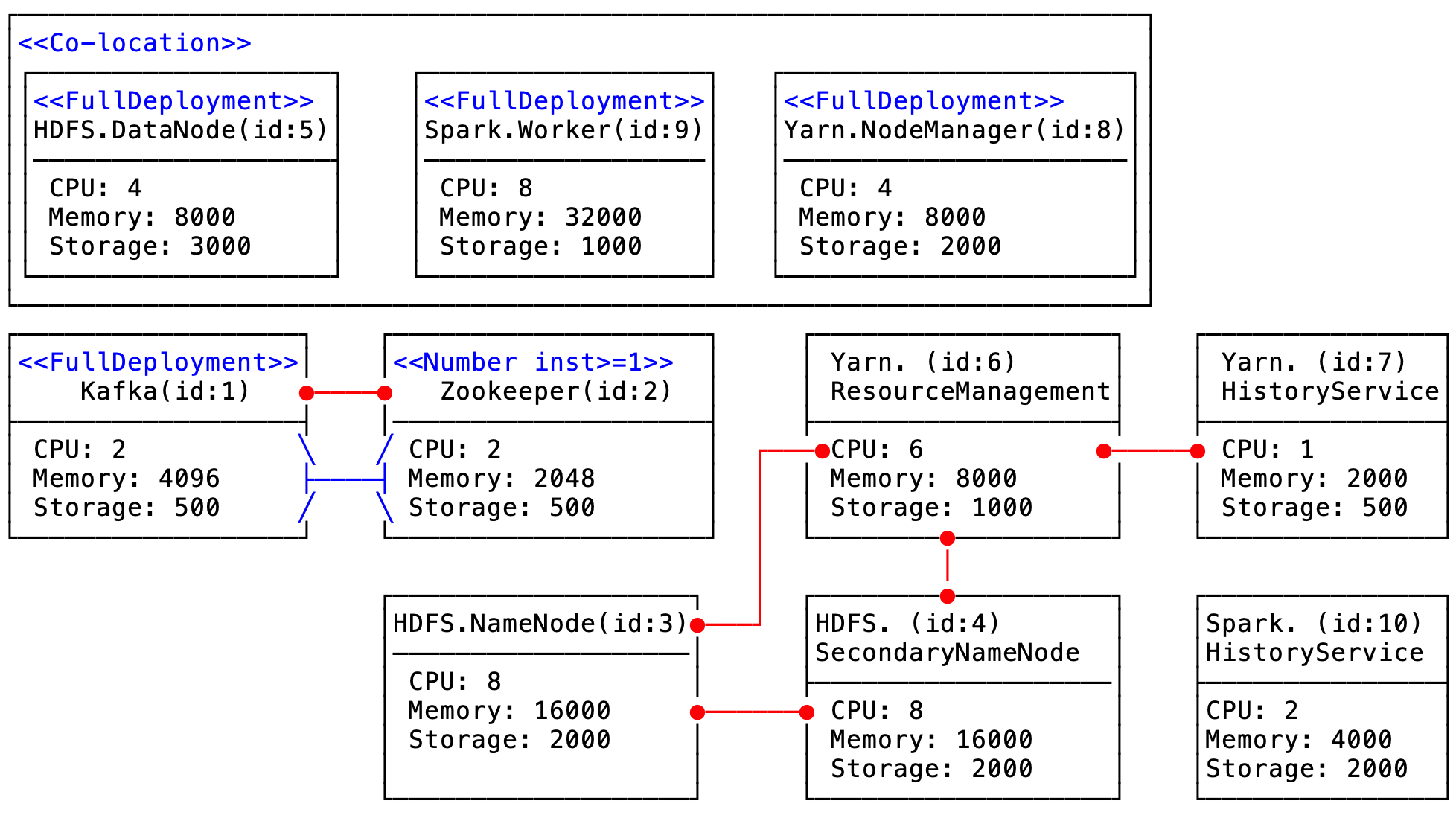}
	\caption{Oryx2 Application}
	\label{fig:Oryx2}
\end{figure}

\emph{Oryx2} application (Figure~\ref{fig:Oryx2}) is a realization of the lambda architecture, featuring speed, batch, and serving tiers, with a focus on applying machine learning models in data analysis, and deploys the latest technologies such as Apache Spark\footnote{\url{https://spark.apache.org/}} and Apache Kafka\footnote{\url{https://kafka.apache.org/}}. It has a significant number of components interacting with each other and is highly used in practical applications. It consists of several components which can be distributed over thousands of VMs in the case of a full deployment.
The main goal of Oryx2 is to take incoming data and use them to create and instantiate predictive models for various use-cases, e.g. movie recommendation.
It is comprised of several technologies. Both the batch and serving layer are based on Apache Spark which in turn uses both Apache Yarn\footnote{\url{http://hadoop.apache.org/}} for scheduling and Apache HDFS as a distributed file system. For a processing pipeline Oryx2 uses Apache Kafka with at least two topics; one for incoming data and one for model update. Apache Zookeeper\footnote{\url{https://zookeeper.apache.org/}} is used by Kafka for broker coordination. All of the aforementioned technologies have subservices with a minimum system requirement and recommended deployment as of Figure~\ref{fig:Oryx2}.
The constraints corresponding to the interactions between the components are described in the following.
\begin{inparaenum}[\itshape (i)\upshape]
\item Components HDFS.DataNode and Spark.Worker must the deployed on the same VM  (\emph{\textbf{Co-location}}). In this scenario, we also collocated Yarn.NodeManager because we used Yarn as a scheduler for Spark jobs.  
\item Components Kafka and Zookeeper, HDFS.NameNode and HDFS.SecondaryNameNode, YARN.ResourceManagement and HDFS.NameNode, HDFS.SecondaryNameNode, YARN.Histo\-ryService are, respectively, in \emph{\textbf{conflict}}, that is, they must not be placed on the same VM.
\item Components HDFS.DataNode, YARN.NodeManager and Spark.Worker must be deployed on all VMs except those hosting conflicting components  (\emph{\textbf{Full Deployment}}).
\item In our deployment, we consider that for one instance of Kafka there must be deployed exactly 2 instances of Zookeeper  (\emph{\textbf{Require-Provide}} constraint). There can be situations, however, when more Zookeeper instances are deployed for higher resilience.
\item A single instance of YARN.HistoryService, respectively Spark.HistoryService should be deployed (\emph{\textbf{Deployment with bounded number of instances}}  constraint, in particular \emph{\textbf{equal bound}}).
\end{inparaenum}

%% file: problem.tex
\section{The Problem}\label{sec:problem}
To describe the problem in a formal way, we consider a set of $N$ interacting components, \mbox{$C=\{C_1,\ldots, C_N\}$}, to be assigned to a set of $M$ virtual machines, $V=\{V_1, \ldots, V_M\}$. Each component $C_i$ is characterized by a set of requirements concerning the hardware resources. Each virtual machine, $V_k$, is characterized by a \emph{type}, which is comprised by hardware/software characteristics and leasing price. There are also \emph{structural constraints} describing the interactions between components (see Section~\ref{sec:StructuralConstraints}). The problem is to find: 
\begin{enumerate}
	\item assignment matrix $a$ with binary entries $a_{ik}\in \{0,1\}$ for $i=\overline{1,N}$, $k=\overline{1,M}$, which are interpreted as follows: 
	\begin{equation*} \label{eq:asignMat}
	a_{ik}=
	\left\{\begin{array}{ll}
	1 & \hbox{if }C_{i} \hbox{ is assigned to } V_k\\  
	0 & \hbox{if }C_{i} \hbox{ is not assigned to } V_k.
	\end{array}
	\right.
	\end{equation*}
	\item the type selection vector $\emph{t}$ with integer entries $\emph{t}_k$ for $k=\overline{1,M}$, representing the type (from a predefined set) of each VM leased.
\end{enumerate}
such that:
\begin{inparaenum}[\itshape (i)\upshape]
	\item the structural constraints,
	\item the hardware requirements  (capacity constraints) of all components are satisfied and 
	\item the purchasing/ leasing price is minimized. 
\end{inparaenum}

For instance, in the case of a Secure Web Container Service (Section \ref{sec:SecWebContainer}), the problem corresponding to $5$ components and to a prior estimation of the number of VMs equal to $6$, a solution can be~\eqref{eq:a}, respectively~\eqref{eq:vmType}.
\begin{equation}\label{eq:a}
a=\begin{pmatrix}
0 & 0 & 0 & 0 & 1 & 0\\ 
0 & 0 & 1 & 1 & 0 & 0\\ 
0& 1 & 0 & 0 & 0 & 0\\ 
1& 0 & 0 & 0 & 0 & 0\\ 
0 & 1 & 1 & 1 & 0 & 0
\end{pmatrix}
\end{equation}

\begin{equation}\label{eq:vmType}
\emph{t} = [12, 12, 13, 13, 15, 0].
\end{equation}

The \emph{structural constraints} are \emph{application-specific} (see Section~\ref{sec:StructuralConstraints}) and derived in accordance with the analysis of the case studies from Section~\ref{sec:casestudies}. \emph{General constraints} (see Section~\ref{sec:GeneralConstraints}) are always considered in the formalization and are related to the: 
\begin{inparaenum}[\itshape (i)\upshape]
	\item \emph{basic allocation} rules, 
	\item \emph{occupancy} criteria,
	\item hardware \emph{capacity} of the VM offers,
	\item \emph{link} between the VM offers and the components hardware/software requirements.
\end{inparaenum}

The problem to solve can be stated as a COP as follows:


\[
\begin{tabular}{rll}
\emph{\textbf{Minimize}} & $\sum\limits_{k=1}^{M} v_k \cdot \emph{p}_{k}$ & \\
\emph{\textbf{Subject to}} & $a_{ik} \in \{0, 1\} $, $v_{k} \in \{0, 1\} $       & \\
                             & $t_{k}\in \text{predefined finite set of natural numbers}$\footnotemark& $\forall i=\overline{1, N}$, $\forall k=\overline{1, M}$\\
\emph{\textbf{General}}      & 
\textbf{\emph{constraints}}&\\
\emph{Basic allocation}& $\sum\limits_{k=1}^M a_{i k} \ge 1$ &  $\forall i=\overline{1,N}$\\
\emph{Occupancy}& $\sum\limits_{i=1}^N a_{ik} \ge 1 \Rightarrow v_k = 1$ & $\forall k=\overline{1,M}$ \\
\emph{Capacity}& $\sum\limits_{i=1}^{N}{a_{ik}} \cdot \emph{R}_{i}^{h} \le F^{h}_{t_k}$ & $\forall k=\overline{1, M}, \forall h=\overline{1, H}$ \\
\emph{Link} & $v_k \!\!=\!\! 1  \wedge t_{k}\!\! = \!\! o \Rightarrow \bigwedge\limits_{h=1}^{H} \left(r^{h}_{k} \!\! = \! \!\emph{F}_{t_k}^{h}\right)  \wedge p_{k}\!\! = \! \!P_{t_k}$ & $\forall o=\overline{1,O}, \ O \in \mathbb{N}^*$\\
           & $\sum_{i=1}^{N}{a_{ik}} = 0  \Rightarrow \emph{t}_{k} = 0$ & $\forall k=\overline{1,M}$ \\
\emph{\textbf{Application-}}& \emph{\textbf{specific}} \emph{\textbf{constraints}}& \\
\emph{Conflicts}& $a_{i k}+a_{j k}\leq 1$            &$\forall k=\overline{1,M}$, $\forall(i,j) \ \mathcal{R}_{ij}=1$\\
\emph{Co-location}& $a_{i k} = a_{j k}   $           & $\forall k=\overline{1,M}$, $\forall(i,j) \ \mathcal{D}_{ij}=1$ \\
\emph{Exclusive }                     & \emph{deployment} &\\
                   & $\mathcal{H}\left(\sum\limits_{k=1}^M a_{i_{1} k}\right) + ... +\mathcal{H}\left(\sum\limits_{k=1}^M a_{i_{q} k}\right)=1$& for fixed $q\in\{1, ..., N\}$\\
\emph{ }  & &  $\mathcal{H}(u) =  \begin{dcases} 
1 & u > 0 \\[-0.3cm]
0 & u = 0 
\end{dcases}$\\
\emph{Require-}&\emph{Provide} & \\
&$n_{ij}\sum\limits_{k=1}^{M}a_{i k} \leq m_{ij}\sum\limits_{k=1}^{M}a_{j k}$ & $\forall (i,j) \mathcal{Q}_{ij}(n_{ij},m_{ij}) = 1$\\
& $0 \le n\sum\limits_{k=1}^{M}a_{j k} - \sum\limits_{k=1}^{M}a_{i k} < n$ & $n, n_{ij}, m_{ij} \in \mathbb{N^*}$\\
\emph{Full deployment}&$\sum\limits_{k=1}^M\left(a_{i k}+\mathcal{H}\left(\sum\limits_{j,\mathcal{R}_{ij}=1}a_{jk}\right)\right)=\sum\limits_{k=1}^{M} v_k$&\\
\emph{Deployment with} &\emph{bounded number  of instances}  & \\
                           &$\sum\limits_{i \in  \overline{C}} \sum\limits_{k=1}^{M}a_{i k} \langle \hbox{op}\rangle n$ & $|\overline{C}|\le N$, $\langle \hbox{op} \rangle \!\in \! \{=,\leq,\geq\}, n \!\in\! \mathbb{N}$
\end{tabular}
\]
\footnotetext{As of the definition of a VM type, $t_k$ includes $p_k$. If $t_k=0$ then the machine $k$ is not occupied.}
\noindent where:
\begin{itemize}
	\item $\mathcal{R}_{ij}=1$ if components $i$ and $j$ are in conflict (can not be placed in the same VM);
	\item $\mathcal{D}_{ij}=1$ if components $i$ and $j$ must be co-located (must be placed in the same VM);
	\item $\mathcal{Q}_{ij}(n,m)\!\!=\!\!1$ if $C_i$ requires at least $n$ instances of $C_j$ and $C_j$ can serve at most $m$ instances of $C_i$;
	\item $R_i^h \in \mathbb{N}^{*}$ is the hardware requirement of type $h$ of the component $i$;
	\item $F_{t_k}^{h} \in \mathbb{N}^{*}$ is the hardware characteristic $h$ of the VM of type $t_k$.
\end{itemize}

\subsection{General Constraints}\label{sec:GeneralConstraints}
The \emph{basic allocation rules} specify that each component $C_i$ must be allocated to at least one VM, except those being in \emph{Exclusive Deployment} relation (see below).
$$\sum\limits_{k=1}^M a_{i k} \ge 1 \qquad \forall i=\overline{1,N}$$
\emph{Capacity constraints} specify that the total amount of a certain resource type $h$ required by the components hosted on a particular VM does not overpass the corresponding resource type of a VM offer. 
\begin{equation}\label{eq:hardware}
\sum\limits_{i=1}^{N}{a_{ik}} \cdot R_{i}^{h} \le F^{h}_{t_k} \qquad \forall k=\overline{1, M}, \ \forall h=\overline{1, H}, \ \forall t_k = \overline{1,O}, \ O \in \mathbb{N^*}
\end{equation}

For example, the VM offer of type $t_1 \! = \! 1$, characterized by CPU, memory, storage and price is encoded as $(F_{1}^1, F_{1}^2, F_{1}^3, P_{1}) = (64, \ 976 \text{ GB}, \ 1 \text{ GB}, \ 8.403\text{ USD/h})$, 

In order to have a sound formalization, one also needs to link a type of a VM offer to each of the occupied VMs, $t_k = (r_k^1, r_k^2, ..., r_k^H, p_k)$:
\begin{equation}\label{eq:VMOcc}
v_k \!\!=\!\! 1  \wedge t_{k}\!\! = \!\! o \Rightarrow \bigwedge\limits_{h=1}^{H} \left(r^{h}_{k} \!\! = \! \!\emph{F}_{t_k}^{h}\right)  \wedge p_{k}\!\! = \! \!P_{t_k} \qquad \forall k=\overline{1,M}, \ \forall o = \overline{1,O}, \ O \in \mathbb{N}^*
\end{equation}

Since in our approach $M$ denotes an upper estimation (see Section ~\ref{sec:estimationVMs}) for the number of VMs needed for deployment, estimation which actually might be higher than the optimum. 
Hence, the following constraint is needed:
\begin{equation*}\label{eq:VMNotOcc}
\sum_{i=1}^{N}{a_{ik}} = 0  \Rightarrow \emph{t}_{k} = 0, \quad \forall k=\overline{1,M}
\end{equation*}

Vector $v$ is the binary occupancy vector defined as:
\begin{equation}\label{eq:occVector}
\sum\limits_{i=1}^N a_{ik} \ge 1 \Rightarrow v_k = 1, \qquad \forall k=\overline{1,M}
\end{equation}
\noindent i.e. $v_k$ is $1$ if machine $V_k$ is used and is $0$ otherwise.


\subsection{Application-specific Constraints} \label{sec:StructuralConstraints}
We identified two main types of application-specific constraints regarding the components: those concerning the \emph{interactions} between components (conflict, co-location, exclusive deployment) and those concerning the \emph{number of instances} (require-provide, full deployment, deployment with a bounded number of instances).

\noindent\emph{\textbf{Conflict.}} This case corresponds to situations when there are conflictual components which cannot be deployed on the same VM. 
Considering that all conflicts between components are encoded in a matrix $\mathcal{R}$ (i.e. $\mathcal{R}_{ij}=1$ if $C_i$ and $C_j$ are conflictual components and $\mathcal{R}_{ij}=0$ otherwise), the constraints can be described as a set of linear inequalities: $$a_{i k}+a_{j k}\leq 1, \qquad k=\overline{1,M}, \hbox{ for all } (i,j) \hbox{ such that } \mathcal{R}_{ij}=1.$$ 
It should be noted that this type of constraints usually induces an increase in the number of VMs.

\noindent For example, for the Wordpress application, Varnish component exhibits load balancing features. Hence, it should not be deployed on the same VM with HTTPLoadBalancer or DNSLoadBalancer. Moreover, Varnish and MySQL should not be deployed on the same VM because it is best practice to isolate the DBMS level of an application. Therefore, based on the notations in Figure~\ref{fig:Wordpress}, where each component has an assigned identifier, the corresponding constraints are:
$$a_{5 k} + a_{i k} \leq 1, \quad i \in \{2, 3, 4\}, \quad k=\overline{1,M} $$
\noindent\emph{\textbf{Co-location.}} This means that the components in the collocation relation should be deployed on the same VM. The co-location relation can be stored in a matrix $\mathcal{D}$ (i.e. $\mathcal{D}_{ij}=1$ if $C_i$ and $C_j$ should be collocated and $\mathcal{D}_{ij}=0$ otherwise) and the constraints can be described as a set of equalities: $$a_{i  k} = a_{j k}, \qquad k=\overline{1,M}, \hbox{ for all } (i,j) \hbox{ such that } \mathcal{D}_{ij}=1$$

\noindent The number of VMs needed decreases with the increase of the number of co-located components.

\noindent For example, for the Oryx2 application, components HDFS.DataNode and Spark.Worker must be deployed on the same VM. In this scenario, we also co-located Yarn.NodeManager because we used Yarn as a scheduler for Spark jobs:
\begin{equation*}
a_{ik} = a_{9k}, \quad i \in \{5,8\}, \quad k=\overline{1,M}
\end{equation*}

\noindent\emph{\textbf{Exclusive deployment.}} There are cases when from a set of $q$ components $\{C_{i_1}, C_{i_2}, ..., C_{i_q}\}$ only one should be deployed in a deployment plan. Such a constraint can be described as: $$\mathcal{H}\left(\sum_{k=1}^M a_{i_{1} k}\right) + \mathcal{H}\left(\sum_{k=1}^M a_{i_{2} k}\right) + ... +\mathcal{H}\left(\sum_{k=1}^M a_{i_{q} k}\right)=1,$$
where $\mathcal{H}$ is a function defined as: $\mathcal{H}(u)=1$ if $u>0$ and $\mathcal{H}(u)=0$ if $u=0$.

\noindent For example, for the Wordpress application, only one type of Balancer must be deployed (the Balancer components are HTTPLoadBalancer and DNSLoadBalancer). If HTTPLoadBalancer is deployed, a caching component, in our case Varnish, should also be deployed leading to a different set of conflicts:
$$\mathcal{H}\left(\sum_{k=1}^{M} a_{3k}\right) + \mathcal{H}\left(\sum_{k=1}^{M} a_{4k}\right) = 1 \quad \text{and} \quad  \mathcal{H}\left(\sum_{k=1}^{M} a_{3k}\right) + \mathcal{H} \left(\sum_{k=1}^{M} a_{5k}\right) = 1
$$
\noindent\emph{\textbf{Require-Provide.}} A special case of interaction between components is when one component requires some functionalities offered by other components. Such an interaction induces constraints on the number of instances corresponding to the interacting components as follows: 
\begin{inparaenum}[\itshape (i)\upshape]
	\item $C_i$ requires (consumes) at least $n_{ij}$ instances of $C_j$ and 
	\item $C_j$ can serve (provides) at most $m_{ij}$ instances of $C_i$.
\end{inparaenum}
This can be written as:
\begin{equation}\label{eq:reqprov}
n_{ij}\sum_{k=1}^{M}a_{i k} \leq m_{ij}\sum_{k=1}^{M}a_{j k}, \ n_{ij},m_{ij} \in \mathbb{N}.
\end{equation}
For example, for Wordpress application, the Wordpress component ($C_1$) requires at least three instances of MySQL and MySQL ($C_2$) can serve at most 2 Wordpress instances, leading to the constraint:
$$3 \sum_{k=1}^{M}a_{1 k} \le2\sum_{k=1}^{M}a_{2 k}, \quad k=\overline{1,M}.$$

\noindent A related case is when for each set of $n$ instances of component $C_{i}$ a new instance of $C_{j}$ should be deployed. This can be described as: 
\begin{equation}\label{eq:floor}
0 \le n\sum_{k=1}^{M}a_{j k} - \sum_{k=1}^{M}a_{i k} < n, \ n \in \mathbb{N}
\end{equation}
This constraint cannot be deduced from \eqref{eq:reqprov} because of the following. Taking in \eqref{eq:reqprov} $n_{ij}=1$, we obtain an expression meaning that for $m_{ij}$ instances of $C_j$ one should have at least one instance of $C_i$ (but there can be more). \eqref{eq:floor} is more specific requiring exactly one instance of $C_j$.
\noindent\emph{\textbf{Full deployment.}} 
There can be also cases when a component $C_i$ must be deployed on all leased VMs (except on those which would induce conflicts on components). This can be expressed as: $$\sum_{k=1}^M\left(a_{i k}+\mathcal{H}\left(\sum_{j,\mathcal{R}_{ij}=1}a_{jk}\right)\right)=\sum_{k=1}^M v_k$$ where $\mathcal{R}$ is the conflicts matrix and $\mathcal{H}$ is defined as above.

\noindent For example, for the Oryx2 application, components HDFS.DataNode ($C_5$), YARN.NodeManager ($C_8$) and Spark.Worker ($C_9$) must be deployed on all VMs except those hosting conflicting components. Since $\mathcal{H}\left(\sum\limits_{j,\mathcal{R}_{ij}=1}a_{jk}\right) = 0$, we have $\sum\limits_{k=1}^M a_{i k}=\sum\limits_{k=1}^M v_k$, for $i \in \{5,8,9\}$.

\noindent Note that we do not allow, in the application description, the full deployment of two conflicting components. 

\noindent\emph{\textbf{Deployment with bounded number of instances.}} There are situations when the number of instances corresponding to a set, $\overline{C}$, of deployed components should be equal, greater or less than some values. These types of constraints can be described as follows: 
$$\sum_{i \in \overline{C}} \ \sum_{k=1}^{M}a_{i k} \ \langle \hbox{op}\rangle \ n, \quad \langle \hbox{op} \rangle \in \{=,\leq,\geq\}, \ n \in \mathbb{N}$$

\noindent For example, for the Secure Web Container application, the total amount of instances of components Apache ($C_2$) and Nginx ($C_3$) must be at least $3$ (level of redundancy):
\begin{equation*}
\sum_{k=1}^{M}a_{2k} + \sum_{k=1}^{M}a_{3k} \ge 3
\end{equation*}

%% file: solution.tex
\section{Solving Approach}\label{sec:solution}
Our solving approach is based on the following  methodology:
\begin{inparaenum}[\itshape (i)\upshape]
\item Estimate an upper bound on the number of VMs needed for the application deployment (Section~\ref{sec:estimationVMs}).
\item Analyze the application-specific constraints, in particular co-location and conflicts, and adapt the formalization in such a way that constraints are implicitly satisfied (co-location) or the search space is reduced (conflicts) (Section~\ref{sec:analysisAppSpecConstrs}).
\item Analyze the symmetries of the problem and identify symmetry breaking strategies (Section~\ref{sec:symmetries}).
\item Select the optimization method (Section~\ref{sec:optimizationApproaches}).
\end{inparaenum}

\subsection{Prior Estimation of the Number of Virtual Machines}\label{sec:estimationVMs}
The number of decision variables (elements of the assignment matrix $a$ and of the type selection vector $t$) depends on the number of VMs, $M$, taken into account in the deployment process. However, the optimal number of used VMs is also unknown, thus a prior estimation of a (tight) upper bound is required. In the case of the traditional bin-packing problem, such an upper bound is given by the number of items. In the case of resource allocation class of problems addressed in this paper, since the number of instances per component is also unknown, estimating an upper bound for the number of VMs is not trivial.  In order to estimate $M$ we solve a \emph{surrogate} optimization problem which takes into account only constraints involving number of instances (i.e. \emph{Require-Provide} and  \emph{Deployment with bounded number of instances}) and minimizes the total number of instances:

\begin{tabular}{rll}
\emph{\textbf{Minimize}} & $M=\sum\limits_{i=1}^{N} \nu_i$ & \\
\emph{\textbf{Subject to}} & $\nu_i \in \mathbb{N}^{*},$& $i=\overline{1,N}$ \\
\emph{Require-Provide} & $n_{ij}\nu_i \leq m_{ij}\nu_j$ & $\forall (i,j) \mathcal{Q}_{ij}(n_{ij},m_{ij}) = 1$\\
                                & $0\leq n\nu_j - \nu_i <n$ & \\
\emph{Bounded number of instances}  &	$\sum\limits_{i\in \overline{C}}\nu_i\geq n$ & $|\overline{C}|\le N$			
\end{tabular}

As $M$ is set to the sum of the number of instances estimated by solving the above surrogate problem, it means that it corresponds to the case when each instance will be assigned to a distinct machine which would correspond to the case when all components would be in conflict. Since in real world cases usually not all components are in conflict, it follows that $M$ is an upper bound estimate of the number of virtual machines required to satisfy all application-specific constraints. 
\subsection{Analysis of the Application-specific Constraints}\label{sec:analysisAppSpecConstrs}

Some constraints can be satisfied by a proper encoding of the decision variables (as that referring to the fact that only one instance of a component is deployed on a VM) while other ones can be exploited in order to reduce the search space by redefining the models.
\subsubsection{Co-location-type Constraints} \label{sec:collocation} For instance, the co-location-type constraints can be exploited by combining all co-located components in a hyper-component and redefining the existing constraints. Let us consider that $\overline{C}=\{C_{i_1},C_{i_2},\ldots,C_{i_L}\}$ is a set of components which should be co-located, i.e. for each instance of one of the components deployed on a VM, an instance of all the other components should be deployed on the same machine. Thus all components in $\overline{C}$ will have the same number of deployed instances and the original problem of assigning $N$ components to VMs can be reformulated as the problem of assigning $N-L+1$ components (as $L$ components will be replaced with one hyper-component, $C_{i^*}$). The original constraints involving elements of $\overline{C}$ will be also reorganized as follows:
\begin{description}
\item{\emph{Conflicts.}} All conflict type constraints involving elements $C_{i_l}$ of $\overline C$ will be replaced with one constraint involving $C_{i^*}$, i.e. if $R_{i_lj}=1$ then $R_{i^*j}=1$.
\item{\emph{Exclusive Deployment.}} If an element $C_{i_l}$ of the hyper-component $C_{i^*}$ is in an exclusive deployment relation with a component $C_j$ then $C_j$ will be excluded, i.e. $C_{i^*}$ is preferred.
\item{\emph{Full Deployment.}} If at least one component of the hyper-component $C_{i^*}$ appears in full deployment constraint then such a constraint should be added for $C_{i^*}$.
\item{\emph{Require-provide.}} In all require-provide constraints involving components from $\overline C$ these components will be replaced with the hyper-component $C_{i*}$.
\item{\emph{Capacity related constraints.}} The capacity constraints (e.g. CPUs, memory size, storage) concerning the components from $\overline C$ will be aggregated by sum in unique constraints involving the hyper-component.
\end{description}
\subsubsection{Conflict-type Constraints}\label{sec:conflict}

When there are conflictual components, the conflict graph can be used to identify components which should be placed on different machines. This type of constraints can be further exploited by fixing the values of the decision variables which correspond to some of the conflictual components. More specifically, in a first step, all cliques which exist in the conflict graph are identified, i.e. subsets of components which are fully conflicting, meaning that their instances should be deployed on different VMs. Then the clique $\overline{G}=\{C_{j_1},C_{j_2},\ldots,C_{j_L}\}$ is selected, that is the clique with the largest deployment size, i.e. the largest number of instances ($\nu_{j_1}+\nu_{j_2}+\ldots +\nu_{j_L}$). 
Since the instances of all components belonging to $\overline{G}$ should be assigned to different machines, one can fix values of the assignment variables as follows. Each of the $\nu_{j_q}$ instances of component $C_{j_q}$ in $\overline G$ is assigned sequentially to the first available machine, i.e. the following constraints are explicitly set:
\begin{alignat}{3}
& a_{j_q(l+Q_{q-1})}=1, & &\qquad \hbox{ for } l=\overline{1,\nu_{j_q}} &\label{eq:fv1}\\
& a_{j_ql}=0, & & \qquad \hbox{ for } l=\overline{1,Q_{q-1}} \hbox{ and } l=\overline{Q_{q}\!+\!1,Q_L},  &\label{eq:fv2}
\end{alignat}
\noindent where the value $Q_q=\nu_{j_1}+\nu_{j_2}+\ldots +\nu_{j_q}$ denotes the number of machines already occupied by instances corresponding to components $\{C_{j_1},C_{j_2},\ldots,C_{j_q}\}$.
Since, the values $\nu_{j_q}$ obtained by solving the surrogate problem described in subsection \ref{sec:estimationVMs} might represent over-estimations (e.g. particularly for the components involved in constraints related to bounded number of instances) a conservative approach would be to fix variables corresponding to just one instance for each of the components belonging to the selected clique $\overline G$. 

Based on the observations above, as well as on the analysis of our case studies, we grouped the decision variables into three main categories: 
\begin{inparaenum}[\itshape (i)\upshape]
\item variables with fixed values (as those set above);
\item variables bounded by constraints (the variables for which
their values are determined by solving some of these constraints); 
\item variables free of constraints, for which the values are mainly controlled through the optimization criterion (e.g. components which fit into any offered VM and are not involved in constraints describing the interaction between components). \end{inparaenum}

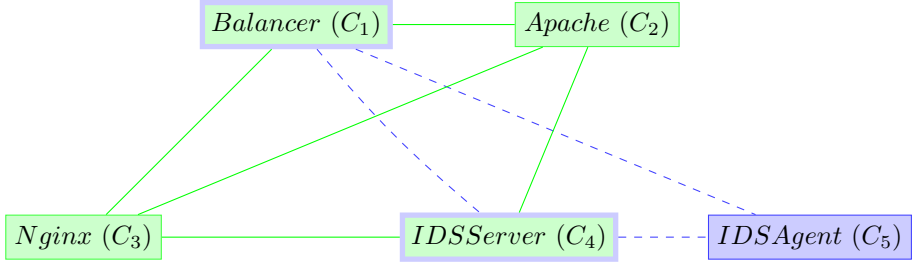
\begin{figure}
    \centering
\begin{tikzpicture}[-,auto,node distance=4cm, scale=0.9]
\tikzstyle{green place}=[rectangle,draw=green!75,fill=green!20]
\tikzstyle{blue place}=[rectangle,draw=blue!75,fill=blue!20]
\tikzstyle{greeny place}=[rectangle,draw=blue!20,fill=green!20,line width=2pt]
  \node[greeny place] (C1)                    {$Balancer \ (C_1)$};
  \node[green place] (C2) [ right of=C1] {$Apache \ (C_2)$};
  \node[green place] (C3) [below left of=C1] {$Nginx \ (C_3)$};
  \node[greeny place] (C4) [below right of=C1] {$IDSServer \ (C_4)$};
  \node[blue place] (C5) [right of=C4]       {$IDSAgent \ (C_5)$};

  \path (C1) edge   [green]           node {} (C2)
             edge    [green]          node {} (C3)
            edge     [green,blue!75,dashed,bend right=5]         node {} (C4)
            edge     [blue!75,dashed]         node {} (C5)
        (C4) edge     [green]         node {} (C2)
             edge     [green]        node {} (C3)
            edge     [blue!75,dashed]  node {}(C5)
        (C2) edge     [green]         node {} (C3);
\end{tikzpicture}
\caption{Secure Web Container conflict graph. The components with green background belong to the clique $\overline{G}$.}
\label{fig:conflictSecure}
\end{figure}

Let us consider the Secure Web Container use case (see Figure~\ref{fig:SecureWebContainer}). The surrogate optimization problem presented in Section~\ref{sec:estimationVMs} estimates a maximum of $6$ VMs for solving the problem. The conflict graph corresponding to the application described in  Figure~\ref{fig:SecureWebContainer}) contains  two cliques: [Balancer, Apache, Nginx, IDSServer] and [Balancer, IDSServer, IDSAgent], the first one having the largest deployment size hence playing the role of $\overline{G}$ (see Figure~\ref{fig:conflictSecure}).
Based on the variable fixing rules described above, most of the decision variables can be fixed, as is illustrated in Table~\ref{tab:fixVarSecWebContainer}, where framed elements correspond to fixed values, and the other elements correspond to variables which are bounded by constraints (e.g. Full deployment in the case of $IDSAgent$, Deployment with bounded number of instances for the set $\{Apache, Nginx\}$). Note that the precise number of instances of $Apache$ and $Nginx$ is not known at the moment of variables fixing; we only know that the sum of their instances should be at least~$3$. Hence only one instance for $Apache$, respectively, $Nginx$ is fixed. In this example, we also explored the fact that $C_1$ is in conflict with $C_5$, although $C_5$ is not part of $\overline{G}$.
\begin{table}[h]
\centering
\caption{Fixing variables in the assignment matrix $a$ of the Secure Web Container use case}
\label{tab:fixVarSecWebContainer}
\begin{tabular}{c|cccccc}
      & $VM_1$        &  $VM_2$        & $VM_3$        & $VM_4$        & $VM_5$ & $VM_6$\\ \hline
$C_1$   & \fbox{1}    &  \fbox{0}    &  \fbox{0}   & \fbox{0}    & 0    & 0\\
$C_2$   & \fbox{0}    &  \fbox{0}    &  \fbox{1}   & \fbox{0}    & 1    & 0 \\
$C_3$   & \fbox{0}    &  \fbox{0}    &  \fbox{0}   & \fbox{1}    & 0    & 0 \\
$C_4$   & \fbox{0}    &  \fbox{1}    &  \fbox{0}   & \fbox{0}    & 0    & 0\\
$C_5$   & \fbox{0}    &  \fbox{0}    &        1   & 1            & 1           & 0\\
\end{tabular}
\end{table}
In this example there are no free of constraints variables, but it would be possible to take into account a component which is not involved in structural constraints (e.g. conflicts, co-location, exclusive or full deployment, require-provide) but can be beneficial from a functional point of view (e.g. a monitoring component). In such a case, different deployment plans having the same cost, but different assignments for the free of constraints component, can be generated. 

\subsection{Symmetries and Symmetry Breakers}\label{sec:symmetries}
Realistic Cloud applications might involve the deployment of a large number of components instances on VMs selected from a large pool of offers. This leads to the necessity of solving optimization problems with a large search space, which, however might contain equivalent solutions. The search space can be limited by reducing the number of decision variables (as is illustrated in Section~\ref{sec:collocation}), by reducing the number of unassigned variables (as is illustrated in Section~\ref{sec:conflict}) or by breaking the symmetries related to the decision variables as discussed in this section.

In the following, symmetries which appear in Cloud deployment problems are described and appropriate symmetry breaking strategies are identified. For a self-contained presentation, we first introduce some theoretical notions. 

\subsubsection{Preliminaries}\label{sec:preliminaries}
A \emph{matrix model} \cite{flener:Reform02} is a constraint program that contains one or more matrices of decision variables. \emph{Symmetries} in constraint satisfaction/optimization problems, in general, and in matrix models, in particular, are a key problem since search can revisit equivalent states many times. In order to deal with symmetries, one must first define what they are. We use the definition from \cite{DBLP:conf/cp/FlenerFHKMPW02}, namely \emph{a symmetry is a bijection on decision variables that preserves solutions and non-solutions}. Two variables are \emph{indistinguishable} if some symmetry interchanges their roles in all solutions and non-solutions. These are \emph{variable symmetries.}
The definition can be extended also to \emph{value symmetries}, i.e. symmetries that permute only the values of variables.

For matrix models, symmetry often occurs because groups of objects within a matrix are \emph{indistinguishable}. This leads to \emph{row/column symmetries}. Two rows/columns are indistinguishable if their variables are pairwise identical due to a row/column symmetry. A matrix model has \emph{row/column symmetry} iff all the rows/columns of one of its matrices are \emph{indistinguishable}.  A matrix model has \emph{partial row/column symmetry} iff strict subset(s) of the rows/columns of one of its matrices are \emph{indistinguishable}. Partial row/column symmetry are more often encountered in Cloud deployment problems, as explained in this section.

Elimination of equivalent states, problem known as \emph{symmetry breaking}, has, most of the times, a positive impact in the computation time of the problem solution process. 
Symmetries can be eliminated by using \emph{symmetry breaking} techniques which can be categorized as follows~\cite[Chapter~10]{DBLP:reference/fai/2}:
\begin{enumerate}
\item \emph{Reformulation} means that the problem is remodeled to eliminate some or all symmetries. It proved to be very efficient method for breaking symmetry, but unfortunately there is no known systematic procedure for performing the remodeling process in general.
\item \emph{Static symmetry breaking} adds the so-called \emph{symmetry breaking constraints} before search starts, hence making some symmetric solutions unacceptable while leaving at least one solution in each symmetric equivalence class.
\item \emph{Dynamic symmetry breaking} removes symmetries dynamically during search, adapting the search procedure appropriately.
\end{enumerate} 
In this paper we use \emph{static symmetry breaking} as detailed in Section \ref{sec:experimentalanalysis}.

A natural way to break symmetry is to order the symmetric objects. To break row/column symmetry, one can simply order the rows/columns lexicographically. 
We say that the \emph{rows/columns in a matrix are lexicographically ordered if each row/column is lexicographically smaller than the previous}. 

Symmetries of Cloud deployment problems can be eliminated in a similar manner, as explained in the following.

\subsubsection{Column Symmetries}\label{sec:colsym}
The \emph{column symmetries} for the Cloud deployment problem from Section \ref{sec:problem} are determined by the decision variables $a$ and $t$. In defining the column symmetries one might exploit the different point of views of the problem. On one hand, the variables $\emph{t}_k$, $\forall k = \overline{1,M}$, representing the type of leased VMs, might be considered indistinguishable, thus exhibit symmetry. At the same time but independently, the columns of the assignment matrix $a$, representing columns, might be considered indistinguishable. These correspond to \emph{full variable symmetry}. Symmetry breakers are based on the idea of ordering the columns, for example:
\begin{description}
\item{(i)} decreasing by the number of components:
$\sum\limits_{i=1}^{N}{a_{ik}}  \ge \sum\limits_{i=1}^{N}{a_{i({k+1})}}, \quad \forall k=\overline{1,N-1}$
\item{(ii)} decreasing by lexicographic order of columns: $a_{\star  k}\succ_{lex} a_{\star (k+1)}$ where $a_{\star k}$ denotes the column~$k$.
\item{(iii)} decreasing by lexicographic order of tuples containing the price and the hardware characteristics of a VM, e.g. CPU number, memory, storage.
\end{description} 

On the other hand, after $\emph{t}_k$, $\forall k = \overline{1,M}$ are assigned values, only the VMs of the same type are indistinguishable. This means that only the columns of the assignment matrix $a$ corresponding to VMs of the same type are indistinguishable. This is \emph{partial symmetry} determined by the fact that we turn \emph{partial value symmetry} for $t$ into \emph{partial variable symmetry for $a$.} Symmetry breakers are also based on the idea of  ordering the columns but with some restrictions, for example:
\begin{description}
\item{(i)} decreasing by the number of components for columns representing VMs of the same type: 
\begin{equation}\label{eq:typeload}
( t_{k} \!=\! t_{k+1}) \Rightarrow \left( \sum\limits_{i=1}^{N}{a_{ik}}  \ge \sum\limits_{i=1}^{N}{a_{i({k+1})}}\right), \quad \forall k=\overline{1,N-1}
\end{equation}
\item{(ii)} decreasing by lexicographic order of columns for columns representing VMs of the same type 
\begin{equation}\label{eq:typelex}
(t_k \!= \! t_{k+1}) \Rightarrow  \left( a_{\star k}\!\succ_{lex} \!a_{\star (k+1)} \right), \text{ where }a_{\star k} \text{ denotes the column }k.
\end{equation}
\end{description} 
\subsubsection{Row Symmetries}\label{sec:rowsym}
The \emph{row symmetries} for the Cloud deployment problem from Section~\ref{sec:problem} are determined by different viewpoints on the assignment matrix $a$. On one hand, the problem has row symmetry if by applying a permutation on the row labels of $a$ (i.e. on the set of components $C_1$, $C_2$, $\ldots$, $C_N$), the assignment matrix corresponds to an equivalent solution (all constraints are satisfied and the value of the optimization criteria is the same). 

The main application-specific constraints which induce row symmetry are that related to \emph{conflicts}, as any two components being in a \emph{conflict} relationship are indistinguishable.
One can break this kind of symmetry by computing the clique with maximal deployment size in which all components are pairwise conflictual hence can not be placed on the same VM. More details were given in Section~\ref{sec:conflict}. 

On the other hand, in the case when all components of an application are identical from the point of view of the hardware requirements and there are no application-specific constraints, the problem has full symmetry with respect to the rows. This means that any of the $N!$ permutations of the assignment matrix row will correspond to an equivalent solution. However, it is rarely this case of Cloud deployment applications. Therefore we are dealing rather with partial row symmetry instead of full row symmetry. 


\subsection{Optimization Approaches}\label{sec:optimizationApproaches}

Optimization problems originated from resource management problems can be tackled by \emph{exact} (\emph{constraints programming}, \emph{mathematical programming}) or \emph{inexact} (\emph{meta-heuristic algorithms}) methods. Inexact methods are highly used in the literature because of their low computational time. However, these methods are suboptimal and there are no theoretical results which allow to estimate how far from the real optimum the solution is. The benefit of exact techniques is that they guarantee the optimal solution, with the disadvantage of higher computational time. 

In this paper we used mathematical programming (MP) and constraint programming, in particular OMT solving. The main difference between these two approaches is the usage of different theoretical basis, namely algebra, respectively logical inferences for solution construction.

\subsubsection{Mathematical Programming}\label{sec:mp}
\emph{Mathematical programming} is a branch in the field of operations research dealing with groups of methods for various optimization problems, particularly linear and quadratic optimization problems. The problem addressed in this paper belongs to the class of integer linear  programming problems for which there are well established solving methods.

In our experiments, we used CPLEX solver~\cite{2016_CPLEX_usermanual}. It is the first commercial linear optimizer on the market to be written in the C programming language started being developed 20 years ago. It implements efficient algorithms solving integer programming, mixed integer programming, and quadratic programming problems. Distinct features of CPLEX, which to the best of our knowledge are not available for OMT solvers, are:
\begin{inparaenum}[\itshape (i)\upshape]
\item it allows to specify bounds on variables at the moment of their declaration; this avoids adding additional logical constraints to the model defining these bounds,
\item it exhibits performance variability which we exploited by ordering the constraints and by using pre-processing parameters (see Section~\ref{sec:SW&HDD-design}). 
\end{inparaenum} 
A drawback which we encountered when using CPLEX to solve the problem  was an explosion of additional variables dynamically generated by the solver, which influences negatively the processing time. 
This happens because CPLEX introduces a new variable for each sum appearing in constraints (e.g. sum of decision variables corresponding to the same column as in Eq. \eqref{eq:reqprov}), even if the same sum appears in several constraints, hence the same variable would have been used. In order to avoid such a variables explosion, the set of occupancy decision variables (i.e. vector $v$ in the problem description) has been explicitly introduced in the CPLEX problem specification.

\subsubsection{Satisfiability and Optimization Modulo Theories}\label{sec:SMTOMT}
SMT solving is an extension of satisfiability (SAT) solving by introducing the possibility of stating constraints in some expressive theories, for example arithmetic, data structures, bit-vector expressions and their valid combinations. The idea behind SMT solving is that, given a formula in a certain theory, this is translated into a propositional formula. This is checked for satisfiability using a SAT solver. If unsatisfiability can not be deduced, then a candidate model (variables assignment) is fed to the theory solver. If this candidate model makes the initial formula true, SAT is returned meaning that the initial formula is true for the respective model. If SAT can not be deduced then the theory solver backtracks trying another candidate model. This process is repeated until no more candidate models are found.

For our problem, we needed a SMT solver which exhibits optimization features, the so-called \emph{Optimization Modulo Theories (OMT) solvers}. There are not too many options in this regard:
\begin{inparaenum}[\itshape (i)\upshape]
	\item OptiMathSAT \cite{10.1007/978-3-319-21690-4_27} uses an inline architecture in which the SMT solver (MathSAT5\footnote{\url{http://mathsat.fbk.eu}}) is
	run only once and its internal SAT solver is modified to handle the search for the optima, 
	\item Symba~\cite{DBLP:conf/popl/LiAKGC14} and $\nu Z$ \cite{DBLP:conf/tacas/BjornerPF15} both are based on an offline architecture in which the SMT solver Z3 \cite{10.1007/978-3-540-78800-3_24} is incrementally called multiple times as a black-box.
\end{inparaenum}
Since Symba is not actively maintained and, in our previous work $\nu Z$ outperformed OptiMathSAT \cite{Erascu_SMT_2019}, in this paper we will use $\nu Z$. We refer to it as $Z3$ being an integrating part of it. It is built on top of the mature SMT solver Z3, it was widely used in various application domains and offers a Python API. It takes as an input SMT formulas (constraints and objectives) which are simplified to Pseudo-Boolean Optimization (PBO) constraints. The PBO solver implements a wide range of methods for simplification and generating conflicting clauses, and compiling them into small sorting circuits. At the heart of the tool lies the dual simplex algorithm which is also used to prune branches. 

OMT is similar to constraint programming methods as they both use logical inferences to construct de solution. Hence, encoding the problem in OMT formalism meant almost a one to one translation of the constraints introduced in Section \ref{sec:problem}.

%% file: experimentalanalysis.tex
\section{Experimental Analysis}\label{sec:experimentalanalysis}
The goal of the experimental analysis is two fold. On one hand, we want to asses the \emph{scalability} of state-of-the-art general MP and OMT tools, namely CPLEX~\cite{2016_CPLEX_usermanual}, respectivelly Z3~\cite{DBLP:conf/tacas/BjornerPF15}, in solving COPs corresponding to the case studies from Section \ref{sec:casestudies}. Tests (see Section \ref{sec:results}) revealed that the naive application of general MP and OMT techniques is not sufficient to solve realistic Cloud deployment applications. Hence, on the other hand, we evaluate the \emph{effectiveness} of various static symmetry breaking techniques in improving the computational time of solving these problems (see Section~\ref{sec:ExperimentalSettings}). The scalability and effectiveness are evaluated from two perspectives: number of VMs offers, respectively number of deployed instances of components.  For \emph{Secure Web Container}, \emph{Secure Billing Email} and \emph{Oryx2} applications, we considered up to 500 VMs offers. Additionally, for the \emph{Wordpress} application, we considered up to $12$ instances of the Wordpress component to be deployed. The set of offers was crawled from the Amazon CPs offers list. 
\subsection{Experimental Settings}\label{sec:ExperimentalSettings}
In this section we present and motivate the plethora of symmetry breakers tested on two types of optimization tools for optimization, as well as the characteristics of the hardware we ran the experiments on.
\subsubsection{Selected Symmetry Breaking Strategies}
Aiming to reduce the search space size, a set of strategies have been selected in order to exploit the particularities of the problem:
\begin{inparaenum}[\itshape (i)\upshape] 
\item the VMs needed for application deployment might have different characteristics; 
\item applications components might be in conflict hence conflict-type constraints can be exploited; 
\item the number of instances to be deployed is unknown.
\end{inparaenum}

Our approach is incremental and experimental: we start with traditional symmetry breakers that have been used for other problems related to bin-packing and combine them with the aim of further search space reduction.

\begin{description}
\item{\emph{Price-based ordering} (PR)}. This strategy aims to break symmetry by ordering  the vector containing the types of used VMs decreasingly by price, i.e. 
$p_k\geq p_{k+1},\   k=\overline{1,M-1}.$ This means that the solution will be characterized by the fact that the columns of the assignment matrix will be ordered decreasingly by the price of the corresponding VMs.
\item{\emph{Lexicographic ordering} (LX)}. This corresponds to the traditional strategy aiming to break column-wise symmetries. The constraints to be added aiming to ensure that two columns, $k$ and $(k+1)$ are in a decreasing lexicographic order, i.e. $a_{*k}\succ_{lex} a_{*(k+1)}$, are:
\begin{equation}\label{eq:lx}
\bigwedge\limits_{l=1}^{i-1} (a_{lk}=a_{l(k+1)}) \Longrightarrow  a_{ik}\ge a_{i(k+1)}, \quad \forall i=\overline{1,N}
\end{equation}
\item \emph{Price-based and lexicographic ordering (PRLX)}. The columns corresponding to VMs with the same price can be considered as indistinguishable, thus the induced symmetries can be broken by ordering lexicographically the corresponding columns.
\begin{alignat*}{3}
& p_{k} \ge p_{k+1}, && \quad \forall k=\overline{1,M-1}\\
& p_{k} = p_{k+1}  \Longrightarrow a_{\star k} \succ_{lex} a_{\star (k+1)} & & 
\end{alignat*}

\item{\emph{Fixed values (FV).}} The search space can be reduced also by fixing the values of some variables starting from the application specific constraints. The strategy included in the experimental analysis is based on the exploitation of the conflict-type constraints as described in Section~\ref{sec:conflict}, i.e. the constraints given by~\eqref{eq:fv1}-\eqref{eq:fv2} have been added to the problem specification.
\item{\emph{Fixed values and price ordering (FVPR).}} 
This strategy uses FV to fix, on separate VMs, the conflicting components and PR to order the VMs. The list of VMs is not globally ordered but it is split in sublists which are ordered. This splitting is based on the structure of the clique with maximal deployment size ($\overline{G}$). More specifically, for each component in $\bar{G}$ the sublist containing the VMs on which its instances are deployed is decreasingly ordered based on price. Finally the VMs which do not contain instances of the components in $\bar{G}$ are decreasingly ordered (see Algorithm~\ref{alg:prvf}).

\item{\emph{Fixed values and lexicographic ordering (FVLX).}} This strategy is similar to FVPR, the only difference being the fact that instead of imposing a price-based order on VMs it is applied a lexicographic order on the corresponding columns of the assignment matrix. More specifically, the lines 12 and 15 in Algorithm \ref{alg:prvf} are replaced with $CL.add(\emph{a}_{\star h} \succ_{lex} \emph{a}_{\star(h+1)} \succ_{lex} \ldots \succ_{lex} \emph{a}_{\star(k-1)})$ and  $CL.add(\emph{a}_{\star k} \succ_{lex} \emph{a}_{*(k+1)} \succ_{lex} \ldots \succ_{lex} \emph{a}_{\star M})$, respectively.

\end{description}
\begin{algorithm}
\begin{algorithmic}[1]
	\STATE Find the clique $\overline{G}$ 
	\STATE $k \leftarrow 1$ \quad /* $k$ - VM index */
	\STATE $CL=\emptyset$  \quad /* CL - constraints list */
	\FOR {each component $C_j \in \overline{G}$}
	\STATE $h \leftarrow k$
	\FOR{each instance of $C_j$}
	\STATE $CL.add(a_{jk} = 1)$
	\FOR{each $t \in \{u\ |\ \mathcal{R}_{ju}=1$\}}
	\STATE $CL.add(a_{tk} = 0)$
	\ENDFOR
	\STATE $k \leftarrow k+1$
	\STATE $CL.add(\emph{p}_{h} \geq \emph{p}_{h+1} \geq \ldots \geq \emph{p}_{k-1})$
	\ENDFOR
	\ENDFOR
	\STATE $CL.add(\emph{p}_{k} \geq \emph{p}_{k+1} \geq \ldots \geq \emph{p}_{M})$
	\RETURN $CL$
\end{algorithmic}
\caption{FVPR Algorithm}
\label{alg:prvf}
\end{algorithm}

Note that Algorithm~\ref{alg:prvf} uses FV strategy, hence it requires the identification of all cliques in the conflict graph. At his aim, we used an implementation of Bron-Kerbosch algorithm available in NetworkX library (\url{https://networkx.github.io}). Finding cliques in a graph is a \textsc{NP}-hard problem, however in our case studies the graph size (given by the number of components) is not very large, so this preprocessing step does not increase significantly the execution time.

It is worth mentioning that the strategies involved in the experimental analysis belong to several classes:
\begin{inparaenum}[\itshape (i)\upshape]
\item PR, LX and PRLX correspond to \emph{column symmetry breakers} as they exploit full and partial symmetries in sets of columns corresponding to groups of VMs with similar characteristics (e.g. same price);
\item FV is a \emph{row symmetry breaker} exploiting the conflict-type constraints;
\item FVPR and FVLX \emph{combine column and row symmetry} by incorporating the advantages of the individual types.
\end{inparaenum}

\subsubsection{Software and Hardware Settings}\label{sec:SW&HDD-design}

In the case of the OMT solver Z3, as background theory, the formalization uses quantifier-free linear arithmetic, in particular quantifier-free linear integer arithmetic. This was chosen based on the results obtained in \cite{Erascu_SMT_2019}. Z3 was used with the default values of the parameters.

In the case of CPLEX, besides the binary decision variables corresponding to the elements of the assignment matrix and the integer decision variables corresponding to the vector containing the types of VMs, the binary variables corresponding to the occupancy vector, $v$, have been explicitly included in the CP problem. This has been done in order to limit the number of variables generated by CPLEX for all constraints specified as logical expressions (as stated in Section \ref{sec:mp}). All CPLEX experiments have been conducted using primal reduction pre-processing (ensured by setting the option {\tt parameters.preprocessing.reduce=1}), as the performance in this case was significantly better than in the other cases (no reduction or primal and dual reduction). 

The implementation, as well as the results, are available at \url{https://github.com/Maneuver-PED/RecommendationEngine}.

All tests in this paper were performed on an Lenovo ThinkCentre with the following configuration: Intel$\textregistered$ Core$\textsuperscript{\texttrademark}$ i7 vPro using CPLEX v12 and Z3 v4.8.7. 

\subsection{Results}\label{sec:results}

Table~\ref{tab:Z3_CPLEX_nosymbreaking} includes the results obtained without using symmetry breaking strategies. The estimated number of VMs reported on the second column has been obtained using the method described in Section \ref{sec:estimationVMs}. The list of offers was crawled from the Amazon site\footnote{\url{https://aws.amazon.com/}}. Each list of VM offers covers the main instance types, for example, \texttt{small}, \texttt{medium}, \texttt{large}. The list of offers can be viewed as a containment hierarchy (i.e. the list of 20 offers is included in the list of 40 offers etc.).

The tables include only those cases for which we obtained a result in a 40 minutes timeframe, as this was the time limit in the SMT-COMP'19\footnote{\url{https://smt-comp.github.io/2019/}}. The missing values ({\tiny $-$}) mean that no solution is returned in this timeframe. 

One can observe that Z3 scales better than CPLEX, however, none of the tools scales for problems involving a large number of components' instances (e.g. at least $4$ instances of the Wordpress component) and a large number of offers (more than a couple of dozens). This is due to the fact that the number $O$ of VM offers influences the number of constraints generated, most notable $H \times O \times M$ constraints of type~\eqref{eq:hardware} and $M \times \emph{O}$ constraints of type~\eqref{eq:VMOcc} are generated. 
\begin{table}[H]
\centering
\tiny
\caption{Scalability tests for Z3 and CPLEX tools. Time values are expressed in seconds}
\label{tab:Z3_CPLEX_nosymbreaking}
\begin{tabular}{c||c|cc|cc|cc|cc}
	\multirow{2}{*}{ \textbf{Problem} } & \multirow{2}{*}{\begin{tabular}[c]{@{}c@{}}\textbf{ estimated}\\\textbf{\#VMs } \end{tabular}} & \multicolumn{2}{c|}{\textbf{\#offers=20 } } & \multicolumn{2}{c|}{\textbf{ \#offers= 40 } } & \multicolumn{2}{c|}{\textbf{ \#offers=250 } } & \multicolumn{2}{c}{\textbf{ \#offers=500 } } \\ 
	\cline{3-10}
	&  & \textbf{Z3}  & \textbf{CPLEX}  & \textbf{Z3}  & \textbf{CPLEX}  & \textbf{Z3}  & \textbf{CPLEX}  & \textbf{Z3}  & \textbf{CPLEX} \\ 
	\hline \hline
	\textbf{Oryx2}  & 11 & \begin{tabular}[c]{@{}c@{}}26.5\end{tabular} & \begin{tabular}[c]{@{}c@{}}\textbf{0.16}\end{tabular} & \begin{tabular}[c]{@{}c@{}}20.15\end{tabular} & \begin{tabular}[c]{@{}c@{}}\textbf{0.42}\end{tabular} & \begin{tabular}[c]{@{}c@{}}\textbf{365.77} \end{tabular} & - & \begin{tabular}[c]{@{}c@{}}\textbf{881.34} \end{tabular} & - \\ 
	\hline
	\begin{tabular}[c]{@{}c@{}} \textbf{Sec. Web} \\ \textbf{Container} \end{tabular} & 6 & \begin{tabular}[c]{@{}c@{}}\textbf{0.34} \end{tabular} & \begin{tabular}[c]{@{}c@{}}5.68 \end{tabular} & \begin{tabular}[c]{@{}c@{}}\textbf{0.85} \end{tabular} & \begin{tabular}[c]{@{}c@{}}101.84\end{tabular} & \begin{tabular}[c]{@{}c@{}}\textbf{11.08} \end{tabular} & - & \begin{tabular}[c]{@{}c@{}}\textbf{27.38} \end{tabular} & - \\ 
	\hline
	\begin{tabular}[c]{@{}c@{}} \textbf{Sec. Billing} \\ \textbf{Email} \end{tabular} & 5 & \begin{tabular}[c]{@{}c@{}}0.21 \end{tabular} & \begin{tabular}[c]{@{}c@{}}\textbf{0.13} \end{tabular} & \begin{tabular}[c]{@{}c@{}}\textbf{0.46} \end{tabular} & \begin{tabular}[c]{@{}c@{}}3.62\end{tabular} & \begin{tabular}[c]{@{}c@{}}\textbf{3.03}\end{tabular} & \begin{tabular}[c]{@{}c@{}}324.8 \end{tabular} & \begin{tabular}[c]{@{}c@{}}\textbf{9.2} \end{tabular} & - \\ 
	\hline
	\begin{tabular}[c]{@{}c@{}}\textbf{Wordpress} \\ \textbf{min\#inst = 3} \end{tabular} & 8 & \begin{tabular}[c]{@{}c@{}}\textbf{2.26} \end{tabular} & \begin{tabular}[c]{@{}c@{}}16.17 \end{tabular} & \begin{tabular}[c]{@{}c@{}}\textbf{7.54}\end{tabular} & - & \begin{tabular}[c]{@{}c@{}}\textbf{223.66} \end{tabular} & - & \begin{tabular}[c]{@{}c@{}}\textbf{547.2} \end{tabular} & - \\ 
	\hline
	\begin{tabular}[c]{@{}c@{}}\textbf{Wordpress} \\ \textbf{min\#inst = 4} \end{tabular} & 10 & \begin{tabular}[c]{@{}c@{}}50.37 \end{tabular} & \begin{tabular}[c]{@{}c@{}}\textbf{6.09} \end{tabular} & \begin{tabular}[c]{@{}c@{}}\textbf{385.17} \end{tabular} & - & - & - & - &  -\\\hline
\end{tabular}
\end{table}
To overcome the lack of scalability issue, we applied the symmetry breaking strategies described in Section \ref{sec:ExperimentalSettings}. The results are presented in Table~\ref{tab:Z3_CPLEX_symbr}.
\setlength{\tabcolsep}{0.15cm}
\begin{sidewaystable}
\caption{Efficiency analysis: time to find an optimal solution (in seconds) in case of applying different symmetry breaking strategies}
\label{tab:Z3_CPLEX_symbr}
\tiny
\centering
\begin{tabular}{c|llllll|llllll|llllll|llllll}
\multirow{2}{*}{\textbf{Problem}} & \multicolumn{6}{c}{\textbf{\#offers=20}} & \multicolumn{6}{c}{\textbf{\#offers=40}} & \multicolumn{6}{c}{\textbf{\#offers=250}} & \multicolumn{6}{c}{\textbf{\#offers=500}} \\
 & PR & LX & FV & PRLX & FVPR & FVLX & PR & LX & FV & PRLX & FVPR & FVLX & PR & LX & FV & PRLX & FVPR & FVLX & PR & LX & FV & PRLX & FVPR & FVLX \\
 \hline\hline
 \multicolumn{17}{c}{Z3 Solver}\\\hline \hline
\textbf{Oryx2} & 0.44 & \textbf{0.33} & 0.49 & 0.52 & 0.38 & 0.45 & 1.21 & 1.27 & 1.34 & 2.42 & \textbf{1.04} & 1.47 & 3.89 & 10.75 & 16.53 & 5.99 & \textbf{2.34} & 6.97 & 10.08 & 39.02 & 84.41 & 19.41 & \textbf{9.83} & 33.70 \\ \hline
\textbf{\begin{tabular}[c]{@{}c@{}}Sec. Billing \\ Email\end{tabular}} & 0.12 & 0.16 & 0.09 & 0.29 & \textbf{0.08} & 0.20 & 0.30 & 0.44 & \textbf{0.27} & 0.52 & \textbf{0.27} & 0.38 & 0.80 & 2.61 & 0.76 & 1.58 & \textbf{0.68} & 1.68 & 2.30 & 7.71 & 1.92 & 5.95 & \textbf{1.85} & 2.68 \\ \hline
\textbf{\begin{tabular}[c]{@{}c@{}}Sec. Web \\ Container\end{tabular}} & 0.13 & 0.16 & 0.12 & 0.21 & \textbf{0.11} & 0.20 & 0.41 & 0.60 & \textbf{0.37} & 0.62 & 0.39 & 0.53 & 0.99 & 3.35 & \textbf{1.52} & 1.65 & 1.54 & 3.86 & 3.24 & 6.70 & \textbf{2.37} & 4.40 & 3.48 & 4.69 \\ \hline
\textbf{\begin{tabular}[c]{@{}c@{}}Wordpress\\  min\#inst=3\end{tabular}} & 0.27 & 0.32 & 0.31 & 0.27 & \textbf{0.23} & 0.44 & \textbf{0.69} & 1.15 & 1.09 & 1.18 & 0.71 & 1.29 & \textbf{1.36} & 7.10 & 12.57 & 5.00 & 2.46 & 6.13 & 4.27 & 29.45 & 24.36 & 10.88 & \textbf{8.05} & 25.49 \\ \hline
\textbf{\begin{tabular}[c]{@{}c@{}}Wordpress \\ min\#inst=4\end{tabular}} & 0.41 & 0.59 & 0.58 & 0.61 & \textbf{0.35} & 0.61 & 1.34 & 2.09 & 2.73 & 1.83 & \textbf{1.11} & 2.63 & 3.92 & 13.65 & 49.84 & 6.64 & \textbf{3.63} & 14.89 & 12.26 & 54.50 & 161.00 & 21.33 & \textbf{9.15} & 37.99 \\ \hline
\textbf{\begin{tabular}[c]{@{}c@{}}Wordpress \\ min\#inst=5\end{tabular}} & 0.60 & 0.88 & 1.11 & 0.71 & \textbf{0.48} & 1.06 & 2.23 & 2.98 & 5.84 & 2.61 & \textbf{1.63} & 2.91 & 12.17 & 18.12 & 167.88 & 8.55 & \textbf{7.26} & 21.46 & 22.86 & 77.51 & 787.07 & 33.53 & \textbf{15.25} & 72.55 \\ \hline
\textbf{\begin{tabular}[c]{@{}c@{}}Wordpress\\  min\#inst=6\end{tabular}} & 0.82 & 1.17 & 1.40 & 1.25 & \textbf{0.49} & 1.16 & 2.40 & 4.50 & 6.11 & 3.11 & \textbf{1.70} & 4.39 & 11.37 & 24.25 & 824.55 & 15.98 & \textbf{6.36} & 45.78 & 25.96 & 117.47 & - & 45.50 & \textbf{17.31} & 59.71 \\ \hline
\textbf{\begin{tabular}[c]{@{}c@{}}Wordpress\\  min\#inst=7\end{tabular}} & 1.15 & 1.68 & 2.32 & 1.36 & \textbf{0.75} & 1.46 & 4.78 & 6.05 & 19.30 & 4.49 & \textbf{2.47} & 7.76 & 23.93 & 40.70 & - & 27.24 & \textbf{8.30} & 43.81 & 46.05 & 140.93 & - & 55.32 & \textbf{28.13} & 92.17 \\ \hline
\textbf{\begin{tabular}[c]{@{}c@{}}Wordpress \\ min\#inst=8\end{tabular}} & 1.95 & 2.56 & 15.25 & 2.23 & \textbf{0.99} & 2.56 & 15.91 & 13.06 & 174.09 & 6.60 & \textbf{4.47} & 15.45 & 139.67 & 107.48 & - & 71.43 & \textbf{28.00} & 161.76 & 260.44 & 302.54 & - & 415.67 & \textbf{79.64} & 303.21 \\ \hline
\textbf{\begin{tabular}[c]{@{}c@{}}Wordpress \\ min\#inst=9\end{tabular}} & 1.97 & 3.34 & 23.96 & 2.63 & \textbf{1.06} & 3.10 & 15.62 & 16.50 & 202.34 & 8.00 & \textbf{4.66} & 17.69 & 273.28 & 111.81 & - & 139.91 & \textbf{33.64} & 167.59 & 375.29 & 396.45 & - & 421.62 & \textbf{87.92} & 253.93 \\ \hline
\textbf{\begin{tabular}[c]{@{}c@{}}Wordpress\\  min\#inst=10\end{tabular}} & 5.95 & 4.30 & 71.71 & 3.24 & \textbf{1.50} & 5.41 & 62.44 & 21.93 & 1008.30 & 12.79 & \textbf{6.06} & 26.24 & 492.09 & 126.14 & - & 127.68 & \textbf{34.33} & 170.95 & - & 467.52 & - & 1187.05 & \textbf{240.36} & 476.49 \\ \hline
\textbf{\begin{tabular}[c]{@{}c@{}}Wordpress \\ min\#inst=11\end{tabular}} & 27.24 & 5.45 & 253.62 & 3.46 & \textbf{1.91} & 5.97 & 216.74 & 28.96 & 2328.61 & 24.74 & \textbf{6.76} & 41.91 & 2078.36 & 155.49 & - & 414.06 & \textbf{70.77} & 239.72 & - & 436.16 & - & - & \textbf{338.74} & 671.25 \\ \hline
\textbf{\begin{tabular}[c]{@{}c@{}}Wordpress\\  min\#inst=12\end{tabular}} & 28.43 & 5.55 & 239.09 & 4.67 & \textbf{1.92} & 7.14 & 225.30 & 32.69 & - & 17.64 & \textbf{8.41} & 33.03 & 1755.92 & 188.86 & - & 414.70 & \textbf{33.32} & 245.59 & - & 690.17 & - & - & \textbf{418.80} & 623.72
\\\hline\hline
\multicolumn{17}{c}{CPLEX Solver}
\\\hline
\textbf{Oryx2} & 0.05 & 14.06 & 0.20 & 1.36 & \textbf{0.03} & 1.23 & 0.14 & 68.22 & 0.19 & 5.45 & \textbf{0.06} & 12.35 & 5.25 & - & 60.08 & 94.82 & \textbf{3.04} & 925.00 & 247.11 & - & - & - & \textbf{27.30} & - \\ \hline
\textbf{\begin{tabular}[c]{@{}c@{}}Sec. Billing\\  Email\end{tabular}} & 0.07 & 1.32 & 0.05 & 0.38 & \textbf{0.06} & 0.11 & 0.35 & 5.03 & 0.23 & 1.48 & \textbf{0.21} & 0.42 & 58.56 & - & 29.22 & 234.97 & \textbf{31.98} & 37.89 & - & - & - & - & \textbf{376.39} & - \\ \hline
\textbf{\begin{tabular}[c]{@{}c@{}}Sec. Web\\  Container\end{tabular}} & 0.31 & 10.53 & 0.15 & 2.40 & 0.12 & \textbf{0.10} & 3.60 & 142.85 & 0.33 & 12.78 & \textbf{0.30} & 0.31 & 347.47 & - & 74.70 & - & \textbf{68.86} & 58.08 & - & - & - & - & \textbf{1316.53} & 1795.03 \\ \hline
\textbf{\begin{tabular}[c]{@{}c@{}}Wordpress\\  min\#inst=3\end{tabular}} & \textbf{0.19} & 85.58 & 0.89 & 0.72 & 0.24 & 2.37 & \textbf{1.55} & - & 42.77 & 7.88 & 1.97 & 39.62 & \textbf{346.77} & - & - & 1588.69 & 1582.11 & - & - & - & - & - & - & - \\ \hline
\textbf{\begin{tabular}[c]{@{}c@{}}Wordpress \\ min\#inst=4\end{tabular}} & \textbf{0.32} & 910.36 & 1.84 & 0.84 & \textbf{0.32} & 4.93 & 4.48 & - & - & 10.27 & \textbf{3.63} & 153.73 & \textbf{1252.09} & - & - & - & - & - & - & - & - & - & - & - \\ \hline
\textbf{\begin{tabular}[c]{@{}c@{}}Wordpress\\  min\#inst=5\end{tabular}} & \textbf{0.50} & - & 4.05 & 4.18 & 0.54 & 18.63 & 7.04 & - & - & 45.68 & \textbf{3.58} & - & - & - & - & - & - & - & - & - & - & - & - & - \\ \hline
\textbf{\begin{tabular}[c]{@{}c@{}}Wordpress \\ min\#inst=6\end{tabular}} & \textbf{0.67} & - & 14.73 & 5.21 & 0.68 & 18.49 & 14.03 & - & - & 36.24 & \textbf{8.76} & - & - & - & - & - & - & - & - & - & - & - & - & - \\ \hline
\textbf{\begin{tabular}[c]{@{}c@{}}Wordpress \\ min\#inst=7\end{tabular}} & \textbf{0.39} & - & 12.05 & 7.75 & 1.92 & 27.22 & 29.01 & - & - & 124.74 & \textbf{25.04} & - & - & - & - & - & - & - & - & - & - & - & - & - \\ \hline
\textbf{\begin{tabular}[c]{@{}c@{}}Wordpress\\  min\#inst=8\end{tabular}} & \textbf{0.54} & - & 1554.52 & 7.53 & 1.77 & 38.41 & 30.07 & - & - & 141.97 & \textbf{23.68} & - & - & - & - & - & - & - & - & - & - & - & - & - \\ \hline
\textbf{\begin{tabular}[c]{@{}c@{}}Wordpress \\ min\#inst=9\end{tabular}} & \textbf{0.77} & - & 25.57 & 7.46 & 2.40 & 39.10 & \textbf{17.16} & - & - & 73.49 & 26.84 & - & - & - & - & - & - & - & - & - & - & - & - & - \\ \hline
\textbf{\begin{tabular}[c]{@{}c@{}}Wordpress\\  min\#inst=10\end{tabular}} & 3.75 & - & 1908.48 & 9.57 & \textbf{1.87} & 58.52 & 74.72 & - & - & 168.08 & \textbf{43.22} & - & - & - & - & - & - & - & - & - & - & - & - & - \\ \hline
\textbf{\begin{tabular}[c]{@{}c@{}}Wordpress \\ min\#inst=11\end{tabular}} & \textbf{2.55} & - & 98.84 & 24.54 & 2.87 & 109.82 & 96.82 & - & - & 320.81 & \textbf{29.20} & - & - & - & - & - & - & - & - & - & - & - & - & - \\ \hline
\textbf{\begin{tabular}[c]{@{}c@{}}Wordpress\\  min\#inst=12\end{tabular}} & 4.70 & - & - & 13.97 & \textbf{2.29} & 63.03 & 239.28 & - & - & 580.61 & \textbf{44.26} & - & - & - & - & - & - & - & - & - & - & - & - & - \\ \hline
\end{tabular}
\end{sidewaystable}

\subsection{Discussion}

From the results reported in Table~\ref{tab:Z3_CPLEX_symbr}, we can draw the following remarks:
\begin{enumerate}
    \item Using appropriate symmetry breakers and the OMT solver Z3 all problem instances were solved within the established timeframe. On the other hand, CPLEX shows lack of scalability, even if for small size problems it leads in a smaller amount of time to solutions.
    \item The best scalability is obtained by combining traditional column-wise symmetry breakers with search space reduction methods which exploit the graph representation associated to the structural constraints specific to each particular application.
\end{enumerate}

Regarding the first remark, one possible explanation of the  poorer scalability of CPLEX is that the number of variables increases significantly when the number of offers increases (see Table~\ref{tab:CPLEXNumberVars}). This is due to the fact that in a preprocessing step, CPLEX translates the original problem into a MP formulation, in particular, all logical implications are rewritten and auxiliary variables are introduced. 

\begin{table}[H]
\tiny
\centering
\caption{Number and types of variables used by CPLEX (explicit decision variables and auxiliary generated variables) - no symmetry breaking case.}
\label{tab:CPLEXNumberVars}
\begin{tabular}{c||ccc|ccc|ccc|ccc}
\textbf{Problem} & \textbf{Binary} & \textbf{Integer} & \textbf{Total} & \textbf{Binary} & \textbf{Integer} & \textbf{Total} &  \textbf{Binary} & \textbf{Integer} & \textbf{Total} & \textbf{Binary} & \textbf{Integer} & \textbf{Total} \\
 & \multicolumn{3}{c|}{\textbf{\#offers=20}} & \multicolumn{3}{c|}{\textbf{\#offers=40}} & \multicolumn{3}{c|}{\textbf{\#offers=250}} & \multicolumn{3}{c}{\textbf{\#offers=500}} \\ \hline \hline
\textbf{Oryx2} & 387 & 63 & \textbf{450} & 644 & 63 & \textbf{707} & 3339 & 63 & \textbf{3402} & 6547 & 63 & \textbf{6610} \\ \hline
\textbf{\begin{tabular}[c]{@{}c@{}}Sec. Billing\\  Email\end{tabular}} & 148 & 29 & \textbf{177} & 265 & 29 & \textbf{294} & 1490 & 29 & \textbf{1519} & 2948 & 29 & \textbf{2977} \\\hline
\textbf{\begin{tabular}[c]{@{}c@{}}Sec. Web\\  Container\end{tabular}} & 178 & 34 & \textbf{212} & 318 & 34 & \textbf{352} & 1788 & 34 & \textbf{1822} & 3538 & 34 & \textbf{3572} \\\hline
\textbf{\begin{tabular}[c]{@{}c@{}}Wordpress\\  min\#inst=3\end{tabular}} & 242 & 46 & \textbf{288} & 429 & 46 & \textbf{475} & 2389 & 46 & \textbf{2435} & 4722 & 46 & \textbf{4768} \\ \hline
\textbf{\begin{tabular}[c]{@{}c@{}}Wordpress\\  min\#inst=4\end{tabular}} & 302 & 57 & \textbf{359} & 535 & 57 & \textbf{592} & 2985 & 57 & \textbf{3042} & 5902 & 57 & \textbf{5959} \\ \hline
\textbf{\begin{tabular}[c]{@{}c@{}}Wordpress\\  min\#inst=5\end{tabular}} & 362 & 69 & \textbf{431} & 642 & 69 & \textbf{711} & 3582 & 69 & \textbf{3651} & 7082 & 69 & \textbf{7151} \\ \hline
\textbf{\begin{tabular}[c]{@{}c@{}}Wordpress\\  min\#inst=6\end{tabular}} & 392 & 75 & \textbf{467} & 695 & 75 & \textbf{770} & 3880 & 75 & \textbf{3955} & 7672 & 75 & \textbf{7747} \\ \hline
\textbf{\begin{tabular}[c]{@{}c@{}}Wordpress\\  min\#inst=7\end{tabular}} & 452 & 86 & \textbf{538} & 802 & 86 & \textbf{888} & 4477 & 86 & \textbf{4563} & 8852 & 86 & \textbf{8938} \\ \hline
\textbf{\begin{tabular}[c]{@{}c@{}}Wordpress\\  min\#inst=8\end{tabular}} & 511 & 98 & \textbf{609} & 908 & 98 & \textbf{1006} & 5073 & 98 & \textbf{5171} & 10031 & 98 & \textbf{10129} \\ \hline
\textbf{\begin{tabular}[c]{@{}c@{}}Wordpress\\  min\#inst=9\end{tabular}} & 541 & 103 & \textbf{644} & 961 & 103 & \textbf{1064} & 5371 & 103 & \textbf{5474} & 10621 & 103 & \textbf{10724} \\ \hline
\textbf{\begin{tabular}[c]{@{}c@{}}Wordpress\\  min\#inst=10\end{tabular}} & 601 & 115 & \textbf{716} & 1068 & 115 & \textbf{1183} & 5968 & 115 & \textbf{6083} & 11801 & 115 & \textbf{11916} \\ \hline
\textbf{\begin{tabular}[c]{@{}c@{}}Wordpress\\  min\#inst=11\end{tabular}} & 661 & 126 & \textbf{787} & 1174 & 126 & \textbf{1300} & 6564 & 126 & \textbf{6690} & 13014 & 126 & \textbf{13140} \\ \hline
\textbf{\begin{tabular}[c]{@{}c@{}}Wordpress\\  min\#inst=12\end{tabular}} & 691 & 132 & \textbf{823} & 1227 & 132 & \textbf{1359} & 6862 & 132 & \textbf{6994} & 13621 & 132 & \textbf{13753} \\ \hline
\end{tabular}
\end{table}

Regarding the second remark, first we notice that, in the case when Z3 solver is used, out of the single-criterion strategies, the LX one leads to significantly better results than PR and FV strategies. This is particularly true in the case of large size problems (i.e. many components and many offers, as is the case of Wordpress). However, since LX strategy leads to many implication type constraints (Eq. \ref{eq:lx}) this has a negative impact in the case of CPLEX solver.

We observe that the FV strategy scales worse than the other individual strategies in the case of the Wordpress application with a large number of offers and a high number of Wordpress instances. 
On the other hand, as it is expected, FV strategy is effective for problems for which the number of pairwise conflicting components is large and a significant reduction in the number of variables can be obtained (e.g. Secure Billing Email, Secure Web Container - see Table \ref{tab:NumberFixedVals}).  In these cases combining FV with other strategies (e.g. FVPR, FVLX) does not bring a benefit because the search space exhibits none or very few symmetries after applying FV and the additional symmetry breaking constraints only increases the computational burden. 

As column-wise symmetry breaker, the PR strategy has the disadvantage of not being able to break all symmetries (as the VMs used in the deployment might have the same price) but the advantage of leading to simpler, thus less computationally costly, symmetry breaking constraints. Therefore, in the case of CPLEX solver the PR strategy leads to better results than LX strategy. On the other hand, in the case of Z3 solver, LX scales better than PR because it turns out that only few types of VMs are typically used for deployment, hence ordering based on price does not break too many symmetries. 

Out of the three combined strategies (PRLX, FVPR, FVLX), the one leading consistently to the best results is FVPR, especially in the case of large size problem instances.  Even in the cases when CPLEX solver has not reached a solution in the established timeframe for PR and FV, the combined strategy has obtained a good performance.
This is because fixing values to variables is equivalent to their elimination. 

Although one would expect that FVLX outperforms PRLX as variables are fixed, this is the case only for Wordpress with at least $8$ deployed instances. In fact, the benefit of combining FV and LX, even when compared with LX strategy, can be observed only for the problem instances mentioned above.

\begin{table}[]
\centering
\tiny
\caption{Problems characteristics and ratio of fixed decision variables:  Number of fixed values as determined by FV strategy (columns 1-3); Estimated vs. used number of VMs (columns 4-5); Number of components vs. number of deployed instances (columns 6-7).}
\label{tab:NumberFixedVals}
\begin{tabular}{c||ccc|cc|cc}
\textbf{Problem}        & \begin{tabular}[c]{@{}c@{}}\textbf{\#fixed values} \\ \textbf{of $a$}\end{tabular}  & \begin{tabular}[c]{@{}c@{}}\textbf{Total\#vars} \\ \textbf{of $a$}\end{tabular} & \begin{tabular}[c]{@{}c@{}}\textbf{Ratio} \\ \end{tabular} & \begin{tabular}[c]{@{}c@{}}\textbf{estimated} \\ \textbf{\#VMs} \end{tabular} & \begin{tabular}[c]{@{}c@{}}\textbf{\#occupied} \\ \textbf{VMs} \end{tabular} & \begin{tabular}[c]{@{}c@{}}\textbf{\#components} \\\end{tabular} & \begin{tabular}[c]{@{}c@{}}\textbf{\#deployed} \\ \textbf{instances} \end{tabular}\\ \hline \hline
\textbf{Oryx2}      & 6                       & 110                                  & 5\%  & 11& 6 & 10&26        \\\hline
\begin{tabular}[c]{@{}c@{}}\textbf{Sec. Web} \\ \textbf{Container}\end{tabular} & 18    & 30     & 60\%     &6&5&5&8     \\\hline
\begin{tabular}[c]{@{}c@{}}\textbf{Sec. Billing} \\ \textbf{Email}\end{tabular}      & 12                      & 25                                   & 48\%  
  & 5 & 5 &6 &6 
\\\hline
\begin{tabular}[c]{@{}c@{}}\textbf{Wordpress} \\ \textbf{min\#inst=3}\end{tabular}   & 9                       & 40                                   & 22\% 
& 8 &8 & 5 & 8
\\\hline
\begin{tabular}[c]{@{}c@{}}\textbf{Wordpress} \\ \textbf{min\#inst=4}\end{tabular} & 12                      & 50                                   & 24\%    
&10&10&5&10
\\\hline
\begin{tabular}[c]{@{}c@{}}\textbf{Wordpress} \\ \textbf{min\#inst=5}\end{tabular} & 15                      & 60                                   & 25\% &12&12&5&12         \\\hline
\begin{tabular}[c]{@{}c@{}}\textbf{Wordpress} \\ \textbf{min\#inst=6}\end{tabular} & 18                      & 65                                   & 27\%  &13&13&5&13        \\\hline
\begin{tabular}[c]{@{}c@{}}\textbf{Wordpress} \\ \textbf{min\#inst=7}\end{tabular}& 21                      & 75                                   & 28\% &15&15&5&15         \\\hline
\begin{tabular}[c]{@{}c@{}}\textbf{Wordpress} \\ \textbf{min\#inst=8}\end{tabular}& 24                      & 85                                   & 28\%  &17&17&5&17        \\\hline
\begin{tabular}[c]{@{}c@{}}\textbf{Wordpress} \\ \textbf{min\#inst=9}\end{tabular} & 27                      & 90                                   & 30\% &18&18&5&18         \\\hline
\begin{tabular}[c]{@{}c@{}}\textbf{Wordpress} \\ \textbf{min\#inst=10}\end{tabular}& 30                      & 100                                  & 30\%  &20&20&5&20        \\\hline
\begin{tabular}[c]{@{}c@{}}\textbf{Wordpress} \\ \textbf{min\#inst=11}\end{tabular} & 33                      & 110                                  & 30\% &22&22&5&22         \\\hline
\begin{tabular}[c]{@{}c@{}}\textbf{Wordpress} \\ \textbf{min\#inst=12}\end{tabular} & 36                      & 115                                  & 31\%  &23&23&5&23       \\\hline
\end{tabular}
\end{table}





\smallskip

%% file: relatedwork.tex
\section{Related Work}\label{sec:relWork}
The work presented in this paper was performed in the framework of the project \emph{MANeUveR: MANagement agency for cloUd Resources}, which sought to answer questions like: \emph{Which CPs offer the best infrastructure at a fair budget? I am no Cloud expert then what are the characteristics of the infrastructure which best fit my application?} At the heart of MANeUveR lies a \emph{Recommendation Engine} module. It is based on some preliminary results obtained in the European-funded projects MODAClouds\footnote{\url{http://www.modaclouds.eu/}} and SPECS\footnote{\url{http://www.specs-project.eu/}} and the national-funded project AMICAS\footnote{\url{https://amicas.hpc.uvt.ro/}}~\cite{Gupta2015,erascu_rolcg_2015,DBLP:journals/tsc/CasolaBEMR17}, namely:
\begin{inparaenum}[\itshape (i)\upshape]
	\item the idea of a Recommendation Engine came from MODAClouds, since their decision support system did not use automatic matching between user requirements and services characteristics, a feature which was necessary for application deployment in a multi-cloud environment;
	\item the work on automated acquiring and configuring cloud resources based on Security Service Level Agreements performed in SPECS motivated us to study the scalability of existing optimization methods used to enhance certain performance indicators (e.g. price);
	\item in the AMICAS project, we presented some preliminary experiments on the scalability of a newly developed method combining exact and heuristic methods for \emph{resource management} problems but in a simpler context: only conflict type constraints were considered and the price minimization was not addressed.
\end{inparaenum}

Regarding the methods for solving resource management problems, we focus on \emph{mathematical programming} and \emph{SMT/OMT solving}.

\emph{Mathematical programming} is heavily used for the improvement of Cloud resource management (see \cite{ZHANG201623} for a survey). Some of them use symmetry breaking for speeding-up the solution process. As an example, \cite{7814593}	formalizes the so-called temporal bin packing problem, that is assignment of a set of tasks (items) to a set of machines (bins) under capacity constraints (CPU usage) where items have a lifespan such that the cost of using the bins is minimized. The problem is solved from two perspectives: Mixed Integer Programming (CPLEX\footnote{\url{https://www.ibm.com/analytics/cplex-optimizer}}) and Constraint Programming (Gecode\footnote{\url{https://www.gecode.org}}). The problem approached in \cite{7814593} is also a generalization of bin packing but, unlike the class of problems we address, it does not involve interactions (e.g. conflicts) between tasks. The only symmetry breakers used are those imposing an order over the time for which bins are allocated or on the bins usage.

The usage of \emph{SMT/OMT techniques} in Cloud deployment problems is scarce. Related work includes \cite{DBLP:conf/kbse/CosmoLTZZEA14}, where the authors solve the following problem: given a high-level specification of the desired system (the set of available components together with their requirements i.e. constraints between components) and the maximal amount of VMs that can be leased, place the components on the VMs such that the minimum number of VMs is used and automated deployment of the system in the cloud environment is achieved. At this aim, they construct a toolchain composed of Zephyrus, for the planning phase, and Armonic, for the deployment phase. Relevant for our work is Zephyrus, since its input is translated into a set of constraints over non-negative integers (in the MiniZinc\footnote{\url{http://www.minizinc.org}} constraint modeling language) and uses various constraint programming solvers for finding the solution. Different to our approach, they use a predefined number of VM types with known hardware specifications based on the prior knowledge of the application requirements. Zephyrus can be proven correct and complete: it will always find a configuration that is optimal. Paper \cite{DBLP:conf/kbse/CosmoLTZZEA14} is extended by \cite{DBLP:conf/setta/AbrahamCJKM16}.  The problem solved is that, given the application description (interaction constraints and hardware requirements) and VMs specifications, the aim is to minimize the cost and then the number of VMs leased for hosting the application. They achieve this by using CP solvers (ge-code\footnote{\url{https://www.gecode.org}}, Google OR-tools\footnote{\url{https://developers.google.com/optimization}}, chuffed\footnote{\url{https://github.com/chuffed/chuffed}}) and SMT solvers (Z3). The paper presents various comparisons on how the tools perform on different types of constraints (linear/nonlinear), the conclusion being that the SMT solvers perform better for nonlinear constraints while are being outperformed by CP in the linear case. Regarding the similarities/dissimilarities to our approach, we mention the following.
\begin{inparaenum}[\itshape (i)\upshape]
	\item Zephyrus2 solves deployment optimization problems by translating them into a COP encoded in MiniZinc. By default, it solves the resulting multi-objective optimization
	problems by optimizing the first objective function value (cost) and then optimizing the
	other objective function (number of components) sequentially following their order after substituting the
	previously determined optimal values. 
	This solution has the drawback that the solver has to be restarted. Differently, we have a single optimization problem (the optimization criteria being the cost) since, until now, in our use cases we did not encounter situations when more components than required by users are assigned to VMs. 
	\item In Zephyrus2, the list of offers is limited (only four types of VMs); in the MANeUveR approach this list is dynamically constructed based on the cloud providers offers. 
	\item In Zephyrus2, according to the provided examples, only memory and price requirements are specified in the problem description; however their approach can be easily extended to other type constraints. However, no performance information are available for more complex problem formulations.
\end{inparaenum}

SMT solvers were applied also to other types of \textsc{NP}-hard problems, similar to ours. For example, \cite{8843025} solves the Flexible Job-Shop Scheduling Problem: we are given a number of jobs, each job being defined by a sequence of operations. Each operation can be executed by any machine from a given pool of machines for a predefined execution time. A machine can be used by multiple operations but onlycone operation is allowed to execute on a single machine at a time. The task is to assign operations to machines such that the total time to complete all jobs is minimized, i.e. minimization of the makespan. The paper proposes three formulations of the problem which are handled by Z3 SMT solver. The experimental results revealed the fact that the SMT solver does not scale well for all formulations. As a remark, the constraints involved in the formulations have arithmetic simpler than ours, while the number of machines used is rather small (at most $7$).

Regarding the literature on \emph{symmetry breaking techniques} relevant for our problem, there are two categories of papers:
\begin{inparaenum}[\itshape (i)\upshape]
	\item experimental papers using symmetry breaking techniques to speed-up the solution process but which do not apply a methodology in their usage (e.g. \cite{DBLP:conf/setta/AbrahamCJKM16, 7814593});
	\item theoretical papers developing symmetry breaking techniques for different versions of the bin-packing problem but which are not directly applicable to our problem (e.g. \cite{Anjos2010SymmetryIS,DBLP:conf/cp/FlenerFHKMPW02,DBLP:conf/aaai/ReginR11,Kiziltan01symmetrybreaking,DBLP:journals/constraints/Gervet97,JANS20131132}). What makes our problem challenging from the theoretical point of view is the presence of various types of structural constraints. 
\end{inparaenum} 
For example, \cite{Anjos2010SymmetryIS} develops a theory for symmetry breaking for job scheduling problems characterized by the fact that a job can be assigned to a single machine from a pool of different machines and there are not multiple instances of the same job\footnote{In terms of our problem, this is equivalent to have in matrix $a$ on each row precisely one $1$, the rows are not constraints dependent and the columns can not be assigned the same type.}. This leads to a problem where all symmetries can be eliminated. They attempt to extend it to machine scheduling problem\footnote{In terms of our problem, this is equivalent to have in matrix $a$ no restrictions on the number of $1$ in each row.}, which is closer to our problem. However, in this case it is not possible to break all symmetries, hence they also propose an experimental approach. Differently to us, they used dynamic symmetry breaking technique (orbital branching) which proved to be more effective than static symmetry breaking when integrated in CPLEX. It worth noticing that the machine problem they consider is much simpler than ours. In the theoretical framework developed by \cite{DBLP:conf/cp/FlenerFHKMPW02}, they study how lexicographic ordering the rows/columns of a matrix model, respectively their composition influences the symmetry breaking. They applied it to the problem of balanced incomplete
block design generation, that is an arrangement of $n$ distinct objects into $b$ blocks, such that each block contains exactly $d$ distinct objects, each object occurring in exactly $r$ different blocks, and every two distinct objects occur together in exactly $\lambda$ blocks. The experimental results reveal the fact that the column symmetries are more efficient than the row ones, while lexicographically ordering the rows and columns can break most of the compositions of the row and column symmetries. Note that the problem constraints are simpler than ours. 
Paper \cite{DBLP:conf/aaai/ReginR11} does not consider any optimality fulfillment criteria except the minimal number of bins; same paper proposes various symmetry breaking methods whose evaluation is proposed; we did not find any continuation of their work. Paper \cite{Kiziltan01symmetrybreaking} considers a problem similar to ours, namely the rack configuration problem, proposes two formalizations of it each exhibiting various column symmetries, which are systematically described. Differently to our approach, in order to break them, they propose the combination of these formalizations. The problem is simpler than ours as no interactions between the card inserted in the racks are present. The experimental results were performed in CONJUNTO \cite{DBLP:journals/constraints/Gervet97}, a tool different to MP and SMT/OMT as it uses interval reasoning on finite sets. \cite{JANS20131132} studies the efficiency of different symmetry breaking techniques and their combination in the framework of job grouping problems, i.e. the task of assigning a set of jobs, each with a specific set of tool requirements, to machines with a limited tool capacity in order to minimize the number of machines needed. This problem is simpler than ours as:
\begin{inparaenum}[\itshape (i)\upshape]
\item it minimizes the number  of machines used and not the price, 
\item the requirement of each job is constant, 
\item the machines are identical, and 
\item there is no dependency/interaction between jobs.  
\end{inparaenum}
Symmetry determined by the identical machines is eliminated by reformulating the problem. In our case, this is handled by the symmetry breakers \eqref{eq:typeload}, \eqref{eq:typelex} and their valid combinations. Except, problem reformulation, they propose other symmetry breakers, for example variable reduction and lexicographic ordering constraints, both on rows and columns. Distinctly from our approach, which reduces the number of variables by fixing values to certain variables from $a$, their meaning of variable reduction is that the assignment of a job $i$ to a higher indexed machine $j$ is not allowed. The techniques are tested on an academic dataset using CPLEX with symmetry breaking option activated.

%% file: conclusions.tex

\section{Conclusions}\label{sec:Conclusions}

We proposed several strategies to tackle the scalability issues in the case of optimal deployment of component-based applications in the Cloud. On one hand, this issue was addressed by observing that problem symmetry is the result of the possible presence of identical virtual machines, which is propagated at various levels, e.g. the way components are assigned to VMs. On the other hand, the particularities of each application are exploited, in particular the symmetries determined by the cliques of the components being in conflict. 

The main conclusion of the experimental analysis is that by combining simple symmetry breaking strategies, one can obtain an effective one. More specifically, by combining a variable reduction strategy with a column-wise symmetry breaker, the scalability was achieved for all problem instances when using the OMT solver Z3. However, CPLEX proved to be appropriate only for small size problems (those involving a couple of dozens of VM offers). One reason for that is that the formulation of the Cloud deployment problem is general and does not exploit the particularities of the CPLEX solver. Hence, for better scalability of CPLEX, the Cloud deployment problem from Section \ref{sec:problem} should be reformulated. 

As future work we plan to compare the symmetry breaking strategies developed in this paper with the CPLEX different options for symmetry breaking. Moreover, since the current encoding of VM offers leads to many constraints, another line of research will be to identify new encodings which ensure the reduction of the search space.